\documentclass[11pt]{elsarticle}
\usepackage[left=2.5cm,top=3.5cm,right=2.5cm,bottom=2cm,nohead]{geometry}

\usepackage{xcolor}
\usepackage{graphicx}
\usepackage{booktabs}
\usepackage{multicol}
\usepackage{multirow}
\usepackage{balance}
\usepackage{enumitem}
\usepackage{titlesec}
\usepackage{tcolorbox}
\usepackage{setspace}
\usepackage{hyperref}
\usepackage[caption=false]{subfig}

\newcommand{\sbar}[1]{{\color{darkgray}\rule{\dimexpr 0.6cm * #1 / 100}{5pt}\color{lightgray}\rule{\dimexpr 0.6cm * (100 - #1) / 100}{5pt}}}

\newcommand{\datasetsize}[0]{6,785}
\newcommand{\totalDeprecatedProjects}[0]{248}
\newcommand{\totalActiveProjects}[0]{754}
\newcommand{\trainingsetsize}[0]{1,002}
\newcommand{\testsetsize}[0]{5,783}
\newcommand{\testsetClassifiedAsUnmaintained}[0]{2,856}

\newcommand{\testsetClassifiedAsActive}[0]{2,927}

\newcommand{\totalProjectsUnmaintainedByReadme}[0]{112}

\newcommand{\totalMaintainedProjectsBySurveyAnswers}[0]{20}
\newcommand{\totalFinishedProjectsBySurveyAnswers}[0]{54}
\newcommand{\totalDeprecatedProjectsBySurveyAnswers}[0]{41}
\newcommand{\totalOthersAnswersBySurvey}[0]{18}
\newcommand{\totalTruePositivesAnswersBySurvey}[0]{103}
\newcommand{\totalFalsePositivesAnswersBySurvey}[0]{26}
\newcommand{\totalProjectsWithCommitsInTheLastYear}[0]{77}

\newcommand{\totalSurveyParticipantsWithPublicEmail}{323}
\newcommand{\totalProjectsWithDeprecatedMessageInReadmeOfSurvey}{21}
\newcommand{\totalSurveyAnswersAndReadmeWithUnclearAnswers}[0]{133}
\newcommand{\totalSurveyAnswersAndReadme}[0]{129}
\newcommand{\totalSurveyAnswersByDevelopers}[0]{112}
\newcommand{\modelPrecision}[0]{80}
\newcommand{\modelRecall}[0]{96}

\usepackage{soul}
\usepackage{balance}
\usepackage{changepage}
\usepackage{framed}
\makeatletter
 {\par\unskip\endMakeFramed}
\makeatother

\definecolor{formalshade}{rgb}{0.93,0.93,0.93}

\definecolor{darkblue}{rgb}{0.2, 0.2, 0.2}

\newenvironment{formal}{%
  \def\FrameCommand{%
    \hspace{1pt}%
    {\color{darkblue}\vrule width 2pt}%
    {\color{formalshade}\vrule width 4pt}%
    \colorbox{formalshade}%
  }%
  \MakeFramed{\advance\hsize-\width\FrameRestore}%
  \noindent\hspace{-1pt}% disable indenting first paragraph
  \begin{adjustwidth}{}{7pt}%
  \vspace{2pt}\vspace{2pt}%
}
{%
  \vspace{3pt}\end{adjustwidth}\endMakeFramed%
}

\newcounter{resultcounter}

\newcounter{patterncounter}
{
}

\makeatletter
\g@addto@macro{\UrlBreaks}{\UrlOrds}
\makeatother
\title{Is this GitHub Project Maintained? Measuring the Level of Maintenance Activity of Open-Source Projects}

\begin{document}
\titleformat*{\section}{\large\bfseries}
\titleformat*{\subsection}{\large\bfseries}
	

 \cortext[cor1]{Corresponding author}

 \author[1]{Jailton Coelho\corref{cor1}}
 \ead{jailtoncoelho@dcc.ufmg.br}
 
 \author[1]{Marco Tulio Valente}
% \ead{mtov@dcc.ufmg.br}
 
 \author[1]{Luciano Milen}
% \ead{lucianomilen@dcc.ufmg.br}
 
 \author[2]{Luciana L.~Silva}
% \ead{luciana.lourdes.silva@ifmg.edu.br}
 
 \address[1]{Federal University of Minas Gerais, Brazil}
 \address[2]{Federal Institute of Minas Gerais, Brazil}

\begin{abstract}

\noindent{{\em Context:} GitHub hosts an impressive number of high-quality OSS projects. However, selecting ``the right tool for the job'' is a challenging task, because we do not have precise information about those high-quality projects.
{\em Objective:} In this paper, we propose a data-driven approach to measure the level of maintenance activity of GitHub projects. Our goal is to alert users about the risks of using unmaintained projects and possibly motivate other developers to assume the maintenance of such projects.
{\em Method:} We train machine learning models to define a metric to express the level of maintenance activity of GitHub projects. Next, we analyze the historical evolution of 2,927 active projects in the time frame of one year.
{\em Results:} From 2,927 active projects, 16\% become unmaintained in the interval of one year. We also found that Objective-C projects tend to have lower maintenance activity than projects implemented in other languages. Finally, software tools---such as compilers and editors---have the highest maintenance activity over time.
{\em Conclusions:} A metric about the level of maintenance activity of GitHub projects can help developers to select open source projects.}

\end{abstract}

\onehalfspacing

\maketitle
    
\section{Introduction}
\label{sec:introduction}

Open source projects have an increasing relevance in modern software development~\cite{nadia2016roads}, powering applications in practically every domain. Today, over 80\% of the software produced is composed by open source code and this trend is growing.\footnote{\url{https://www.linuxfoundation.org/blog/chaoss-project-creates-tools-to-analyze-software-development-and-measure-open-source-community-health}} In a recent investigation conducted by Sonatype\footnote{\url{https://www.sonatype.com/2019ssc}}, they report that downloads of npm packages reached 10 billion per week and 21,448 new open source components are releases per day. For this reason, open source code can be viewed as the backbone of the digital infrastructure that runs our society~\citep{nadia2016roads}. Furthermore, the emergence of world-wide code sharing platforms---such as GitHub---is contributing to transform open source development in a competitive market. Indeed, in a recent survey with maintainers, we found that the most common reason for the failure of open source projects is the appearance of a stronger competitor in GitHub~\cite{coelho2017why}.

However, GitHub does not include objective data about project's maintenance activity. Users can access historical data about commits or repository popularity metrics, like number of stars, forks, and watchers. However, based on the values of theses metrics, they should judge by themselves whether a project is under maintenance or not (and therefore whether it is worth to use it). In order to help on this decision, in this paper we propose and evaluate a machine learning approach to measure the level of maintenance activity of GitHub projects. Our goal is to provide a simple and effective metric to alert users about the risks of depending on a given GitHub project. This information can also contribute to attract new maintainers to a project. For example, users of libraries facing the risks of discontinuation can be motivated to assume their maintenance.

Previous work in this area relies on the last commit activity to classify projects as unmaintained or in a similar status. For example, \citet{khondhu2013all} use an one-year inactivity threshold to classify {\em dormant} projects on SourceForge. The same threshold is used in works by \citet{mens2014survivability}, \citet{izquierdo2017empirical}, and in our previous work about the motivations for the failure of open source projects~\cite{coelho2017why}. However, in this paper, we do not rely on such thresholds when investigating unmaintained projects due to three reasons. First, because setting a threshold to characterize unmaintained projects is not trivial. For example, in the mentioned works, this decision is arbitrary and it is not empirically validated. Second, our intention is to detect unmaintained projects as soon as possible; preferably, without having to wait for one year of inactivity (or another threshold). Third, our definition of unmaintained projects does not require a complete absence of commits during a given period; instead, a project is considered unmaintained even when sporadic and few commits happen in a time interval. Stated otherwise, in our view, unmaintained projects do not necessarily need to be dead, deprecated, or archived.

In our previous conference paper~\cite{coelho2018identifying}, we proposed a machine learning model to identify unmaintained GitHub projects, using as features standard metrics provided by GitHub, e.g.,~number of commits, forks, issues, and pull requests. Then, we validated the proposed model with the principal developers of 129 projects, achieving a precision of 80\%. Particularly, in this previous work, we provided answers to three research questions:\\[-.3cm]

\noindent{\em RQ1: What is the precision of the proposed machine learning model according to GitHub developers?} The intention was to check precision in the field, by collecting to the feedback provided by the principal developers of popular GitHub projects.\\[-.3cm]

\noindent{\em RQ2: What is the recall of the proposed machine learning model when identifying unmaintained projects?} Recall is more difficult to compute in the field, because it requires the identification of {\em all} unmaintained projects in GitHub. To circumvent this problem, we compute recall considering only projects that declare in their README they are not under maintenance.\\[-.3cm]

\noindent{\em RQ3: How early does the proposed machine learning model identify unmaintained projects?} As mentioned, the proposed model does not depend on an inactivity interval to classify a project as unmaintained. Therefore, in this third question, we investigate how early we are able to identify unmaintained projects, e.g.,~without having to wait for a full year of commit inactivity.\\[-.3cm]

The present work extends our previous conference paper in four key directions. First, we updated our dataset and computed new data points for each feature using data from 2018. Second, we provide answers for two novel research questions:\\[-.3cm]

\noindent{\em RQ4: How long does a GitHub project survive before become unmaintained?} The goal is to investigate the survival probability over time of the projects classified as unmaintained by the proposed machine learning model. Moreover, we analyze the survival probability of these projects under different perspectives (e.g., organizational or individual account, programming language, and application domain).\\[-.3cm]

\noindent{\em RQ5: How often unmaintained projects follow best OSS contribution practices?} We investigate whether projects classified as unmaintained follow a set of best open source contribution practices, recommended by GitHub, such as presence of contributing guidelines, presence of project's license and use of a continuous integration service.\\[-.3cm]

Third, to provide a historical perspective on the valves of the proposed Level of Maintenance Activity (LMA), we use as baseline the projects classified as active in November 2017 and compute new values in the time frame of one year (2018). Fourth, we implemented a public Chrome extension to inform the maintenance level of a GitHub project.

This paper is organized as follows. In Section~\ref{sec:machine-learning}, we present and evaluate a machine learning model to identify unmaintained projects. 
Section~\ref{sec:validation} validates this model with GitHub developers and  projects that are documented as deprecated.
In Section~\ref{sec:characteristics-of-unmaintained}, we assess the characteristics of projects classified as unmaintained by our model.
Section~\ref{sec:level-of-maintenance-activity} defines and discusses the Level of Maintenance Activity (LMA) metric. Section~\ref{sec:threats} lists threats to validity and Section~\ref{sec:related-work} discusses related work. Section~\ref{sec:conclusion} concludes the paper and outlines further work.

\section{Machine Learning Model}
\label{sec:machine-learning}

In this section, we describe our machine learning approach to identify projects that are no longer under maintenance.

\subsection{Experimental Design}
\label{sec:prediction-study-design}

\noindent{\bf Dataset.} We start with a dataset containing the top-10,000 most starred projects on GitHub (in November, 2017). Stars---GitHub's equivalent for {\em likes} in other social networks---is a common proxy for the popularity of GitHub projects~\cite{borges2018stars, borges2016icsme}. Then, we follow three strategies to discard projects from this initial selection. First, we remove 2,810 repositories that have less than two years from the first to the last commit (because we need historical data to compute the features used by the prediction models). Second, we remove 331 projects with null size, measured in lines of code (typically, these projects are implemented in non-programming languages, like CSS, HTML, etc). Finally, we remove 74 non-software projects, which are identified by searching for the following topics: {\em books} and {\em awesome-lists}. We end up with a list of \datasetsize\ projects. 

Next, we define two subsets of systems: {\em active} and {\em unmaintained}. The active (or under maintenance) group is composed by \totalActiveProjects\ projects that have at least one release in the last month, including well-known projects, like {\sc facebook/react}, {\sc d3/d3}, and {\sc nodejs/node}. Thus, we assume that these projects are active (under maintenance). By contrast, the unmaintained group is composed by \totalDeprecatedProjects\ projects, including 104 projects that were explicitly declared by their principal developers as unmaintained in our previous work~\citep{coelho2017why} and 144 {\em archived} projects. Archiving is a feature provided by GitHub that allows developers to explicitly move their projects to a read-only state. In this state, users cannot create issues, pull requests, or comments, but can still fork or star the projects.

\begin{figure*}[!t]
  \centering
\vspace{-0.8mm}
\subfloat[ref1][Age]{\includegraphics[width=0.30\textwidth]{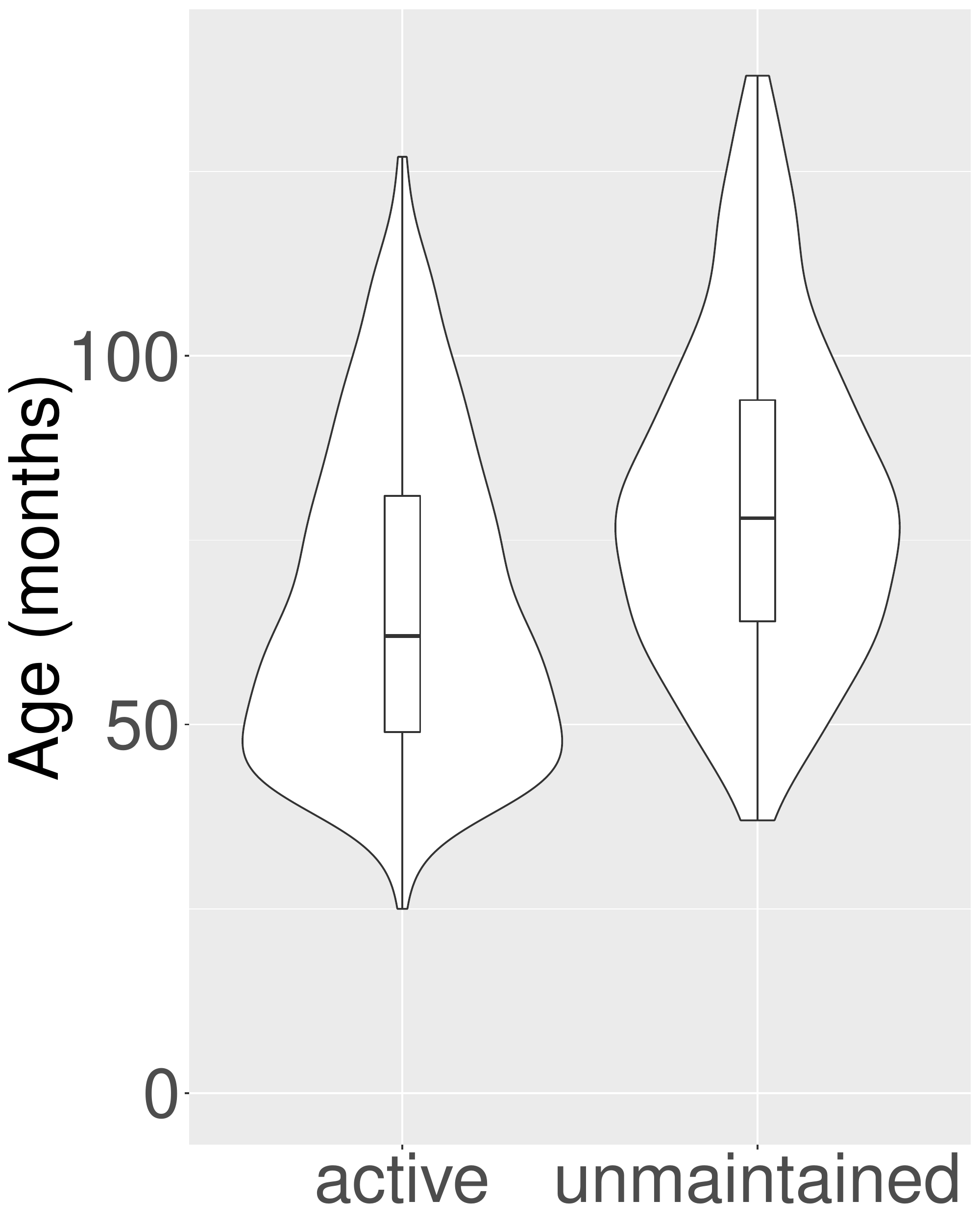}}
\quad
\subfloat[ref2][Forks]{\includegraphics[width=0.30\textwidth]{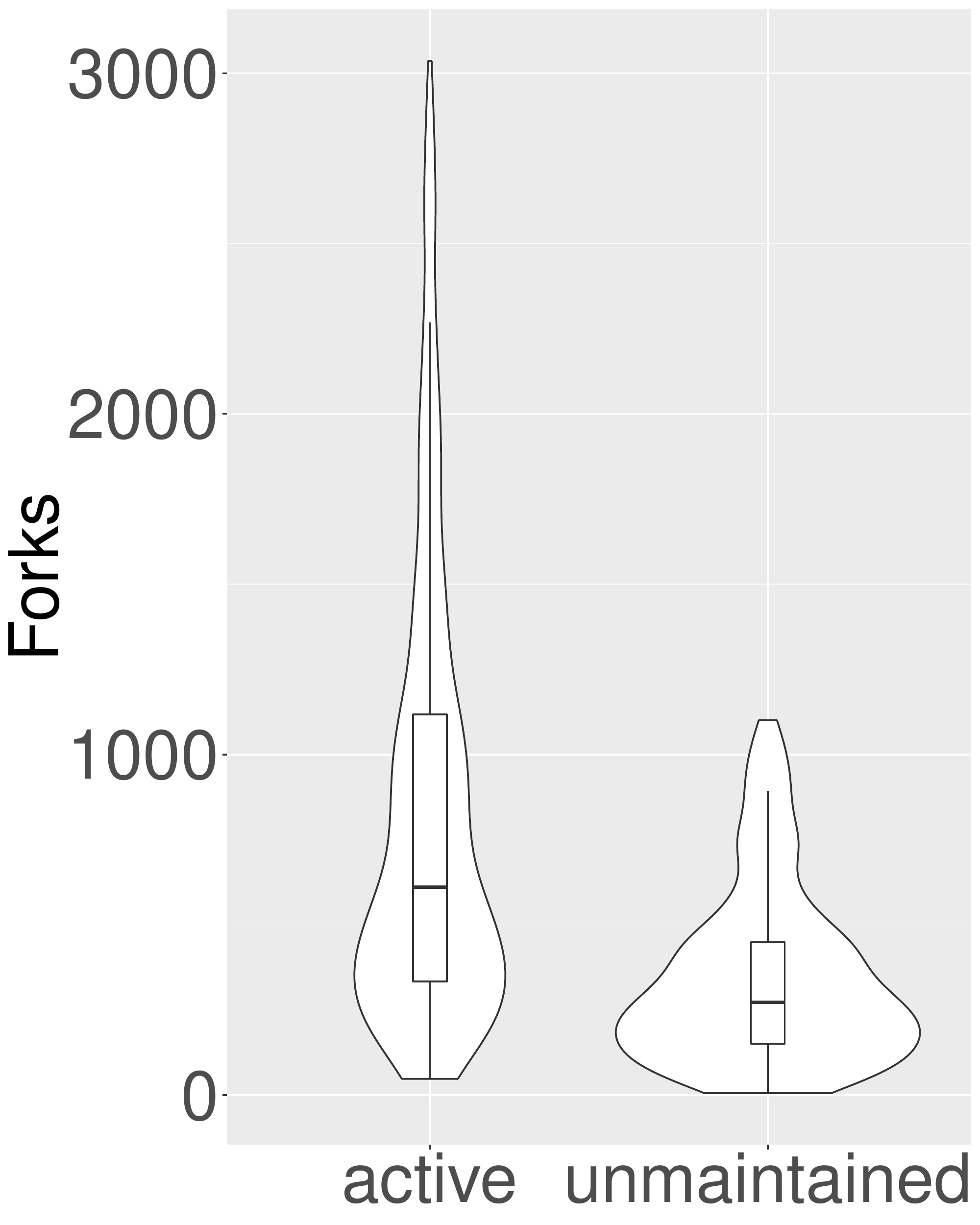}}
\quad
\quad
\quad
\quad
\subfloat[ref3][Commits]{\includegraphics[width=0.30\textwidth]{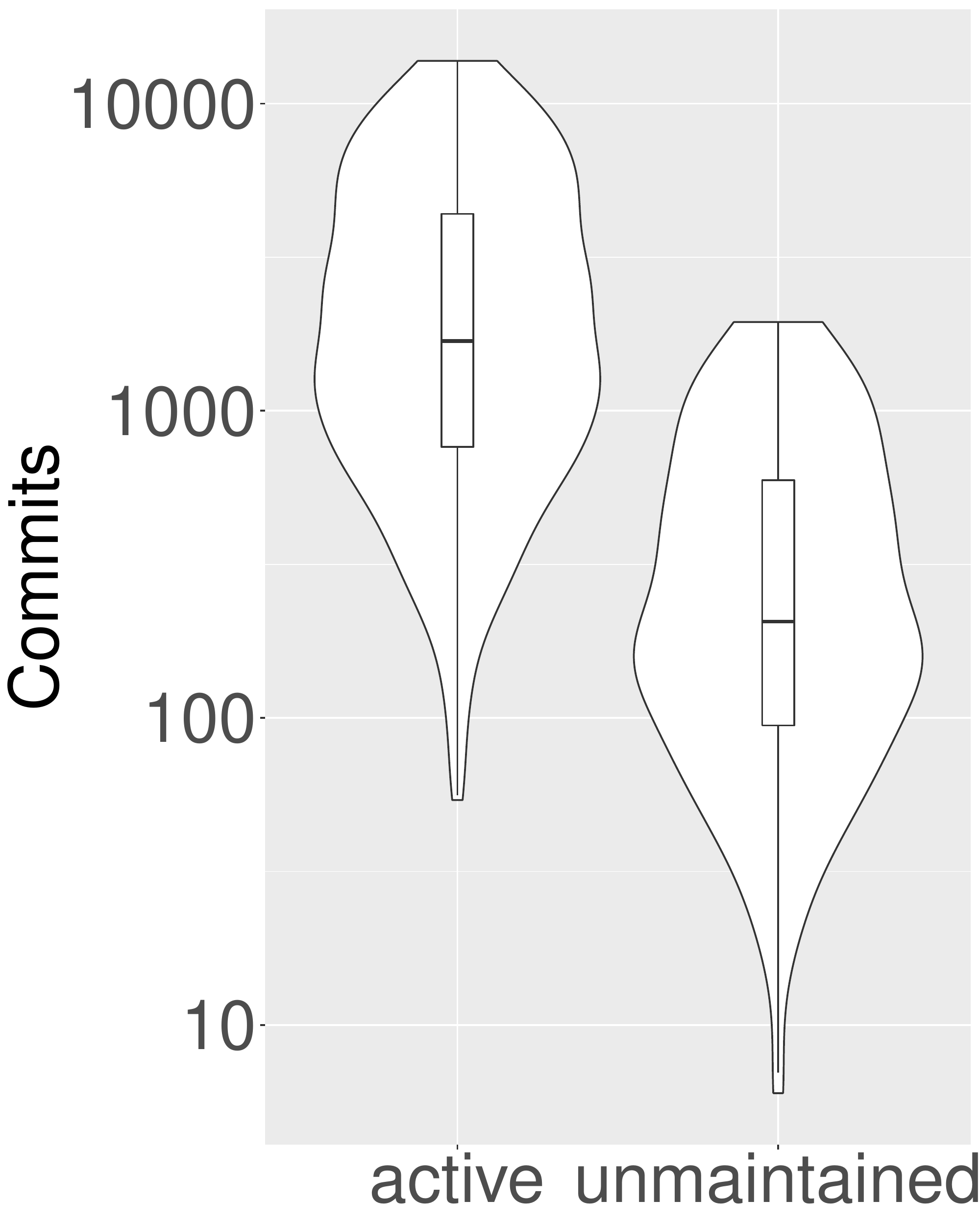}}
\quad
\subfloat[ref4][Stars]{\includegraphics[width=0.30\textwidth]{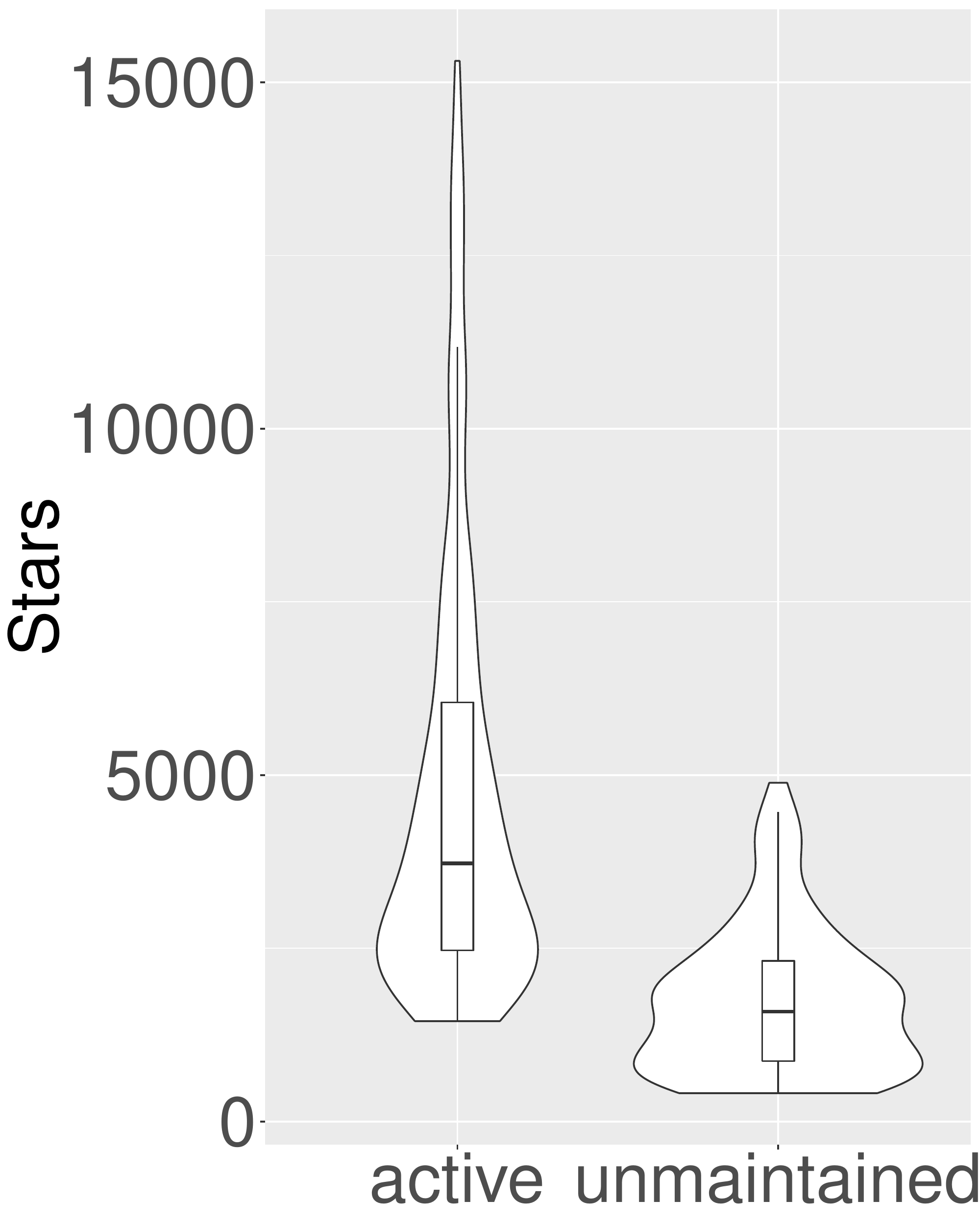}}
\vspace{-3mm}
\caption{Distribution of the (a) age, (b) forks, (c) commits, and (d) stars, without outliers.}
\label{dataset_violinplot_projects_characteristics}
\end{figure*}

\begin{figure*}[!ht]
\centering
\includegraphics[width=7cm]{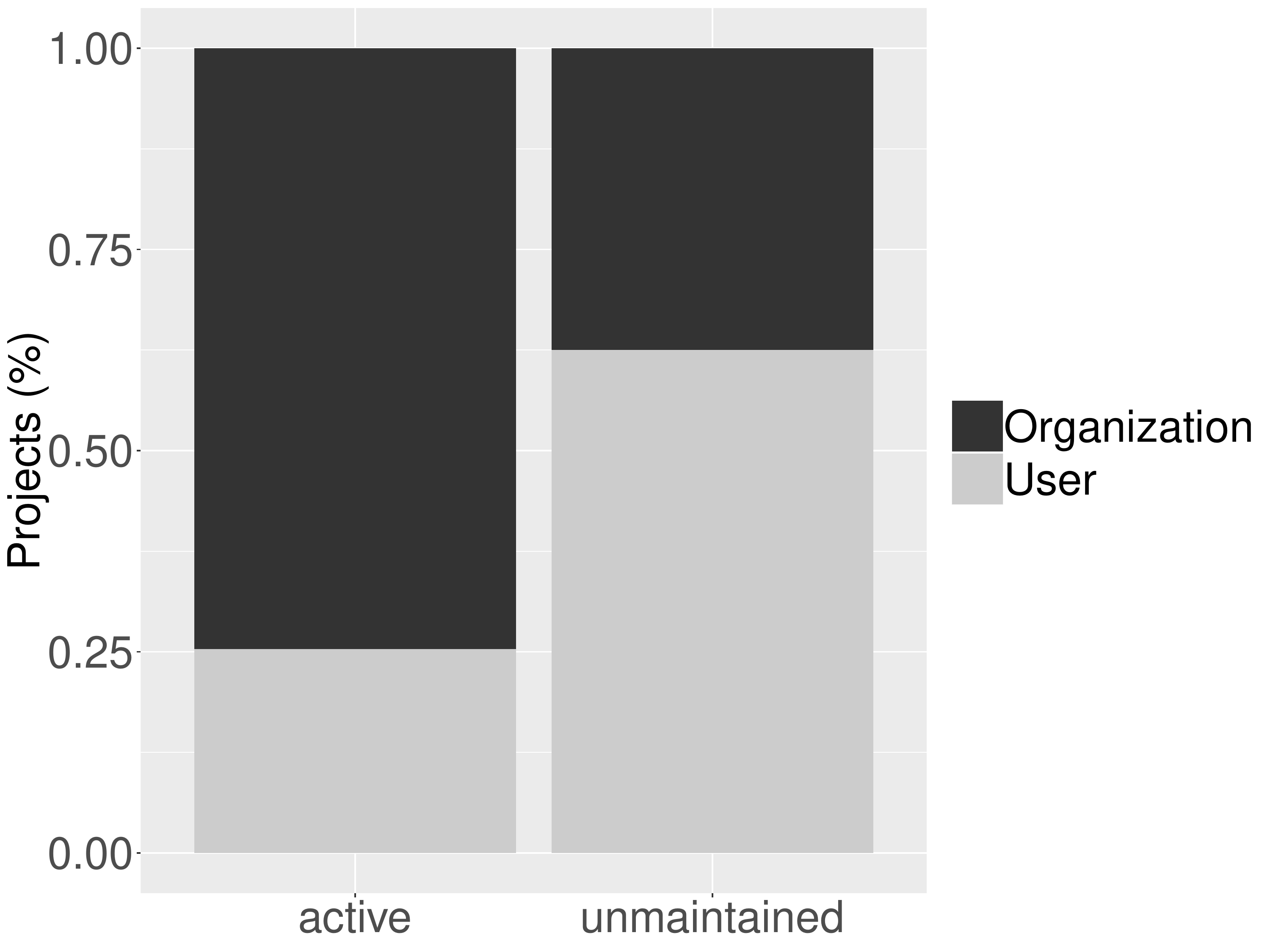}
\caption{Number of projects owned by an individual GitHub user or by an organization.}
\label{fig:barplot_owner}
\end{figure*}

Figure~\ref{dataset_violinplot_projects_characteristics} shows violin plots with the distribution of age (in months), number of forks, number of commits, and number of stars  of the selected repositories. We provide plots for the \totalActiveProjects\ {\em active} projects and for the \totalDeprecatedProjects\ {\em unmaintained}  projects. As we can check, unmaintained projects are older than the active ones (78 vs 62 months, median measures); but they have less forks (299 vs 735), less commits (241 vs 2,136), and less stars (1,714 vs 4,078). In our dataset, active projects are composed by 74\% of organizational projects and 26\% of user projects. By contrast, unmaintained projects consists of 37\% and 63\% of organizational and user projects, respectively (Figure~\ref{fig:barplot_owner}).\\[-.2cm]

\begin{figure}[!ht]
\centering
\includegraphics[width=8.5cm]{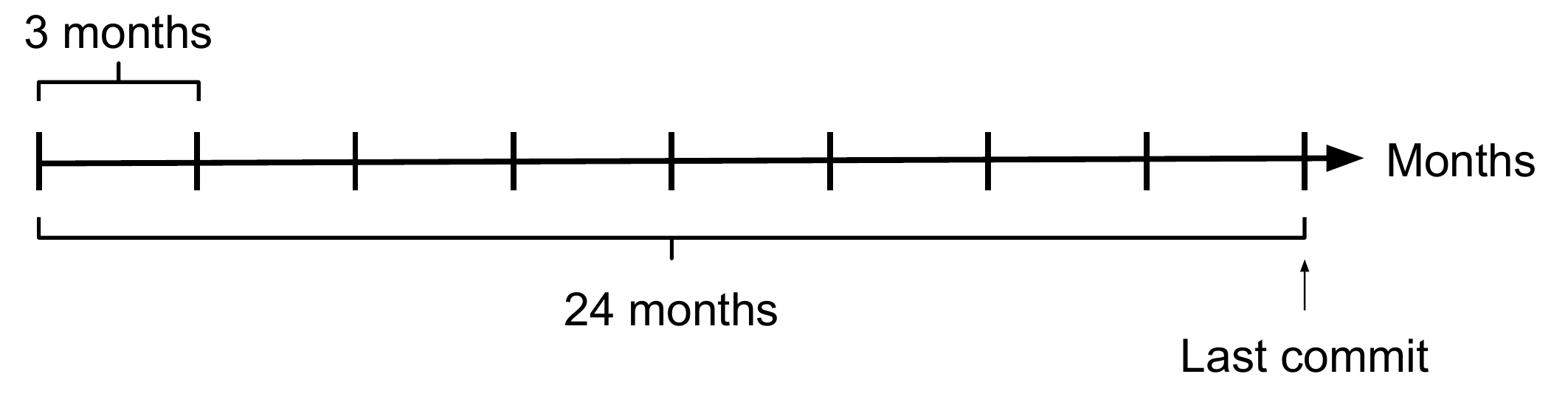}
\caption{Feature collection during 24 months in 3-month intervals.}
\label{fig:intervals}
\end{figure}

\noindent{\bf Features}. Our hypothesis is that a machine learning classifier can identify unmaintained projects by considering features about (1) projects, including number of forks, issues, pull requests, and commits; (2) contributors, including number of new and distinct contributors (the rationale is that maintenance activity might increase by attracting new developers); (3) project owners, including number of projects he/she owns and total number of commits in GitHub (the rationale is that maintenance might be affected when project owners have many projects on GitHub). In total, we consider 13 features, as described in Table~\ref{tab:feature-means}. The feature values are collected using GitHub's official API. However, they do not refer to the whole history of a project, but only to the last $n$ months, counting from the last commit; moreover, we collect each feature in intervals of $m$ months. The goal is to derive temporal series of feature values, which can be used by a machine learning algorithm to infer trends in the project evolution, e.g., an increasing number of opened issues or a decreasing number of commits. Figure~\ref{fig:intervals} illustrates the feature collection process assuming that $n$ is 24 months and that $m$ is 3 months. In this case, for each feature, we collect 8 data points, i.e.,~feature values.\\[-.2cm]

\begin{table*}[!ht]
    \centering
    \caption{Features used to identify unmaintained projects.}
    \small
    \begin{tabular}{ l l l}
        \toprule
        {\bf Dimension} 				& {\bf Feature}			& {\bf Description}			\\ 
        \midrule
        \multirow{9}{*}{\bf Project} 
        &	Forks							& 	Number of forks created by developers	\\
        &	Open issues						& 	Number of issues opened by developers	\\     
        &	Closed issues					& 	Number of issues closed by developers	\\
        &	Open pull requests 				& 	Number of pull requests opened by the developers\\
        &	Closed pull requests			& 	Number of pull requests closed by the developers\\
        &	Merged pull requests			& 	Number of pull requests merged by the developers\\
        &	Commits							& 	Number of commits performed by developers		\\
        & 	Max days without commits		& 	Maximum number of consecutive days without commits	\\
        & 	Max contributions by developer	& 	Number of commits of the developer with most commits  \\
        \midrule
        \multirow{2}{*}{\bf  Contributor}
        &	New contributors				& 	Number of contributors who made their first commit	\\
        & 	Distinct contributors			& 	Number of distinct contributors that committed	\\
        \midrule
        \multirow{2}{*}{\bf  Owner}
        &	Projects created by the owner	& 	Number of projects created by a given owner \\
        &	Number of commits of the owner	& 	Number of commits performed by a given owner  	\\      
        \bottomrule
    \end{tabular}
    \label{tab:feature-means}
\end{table*}

We experiment with different combinations of $n$ and $m$; each one is called a scenario, in this paper. Table~\ref{tab:scenarios} describes the total number of data points extracted for each scenario. This number ranges from 13 data points (scenario with features extracted in a single interval of 6 months) to 104 data points  (scenario with features extracted in intervals of 3 months during 24 months, as in Figure~\ref{fig:intervals}).

\begin{table}[!ht]
    \centering
    \caption{Scenarios used to collect features and train the machine learning models (length and intervals are in months; data points is the total number of data points collected for each scenario).}  
    \begin{tabular}{l r r r r r r r r r r}
     \toprule
	\textbf{Scenario} & \textbf{1} & \textbf{2} & \textbf{3} & \textbf{4} & \textbf{5} & \textbf{6} & \textbf{7} & \textbf{8} & \textbf{9} & \textbf{10} \\
	\hline
    \textbf{Length} 	 & 6  & 6  & 12  & 12  & 12  & 18  & 18  & 24  & 24  & 24 \\
    \textbf{Intervals} 	 & 3  & 6  & 3   & 6   & 12  & 3   & 6   & 3   & 6   & 12 \\
    \textbf{Data points} & 26 & 13 & 52  & 26  & 13  & 78  & 39  & 104 & 52  & 26  \\

    \bottomrule
    \end{tabular}
    \label{tab:scenarios}
\end{table}

\begin{figure}[!ht]
\centering
\includegraphics[width=15cm]{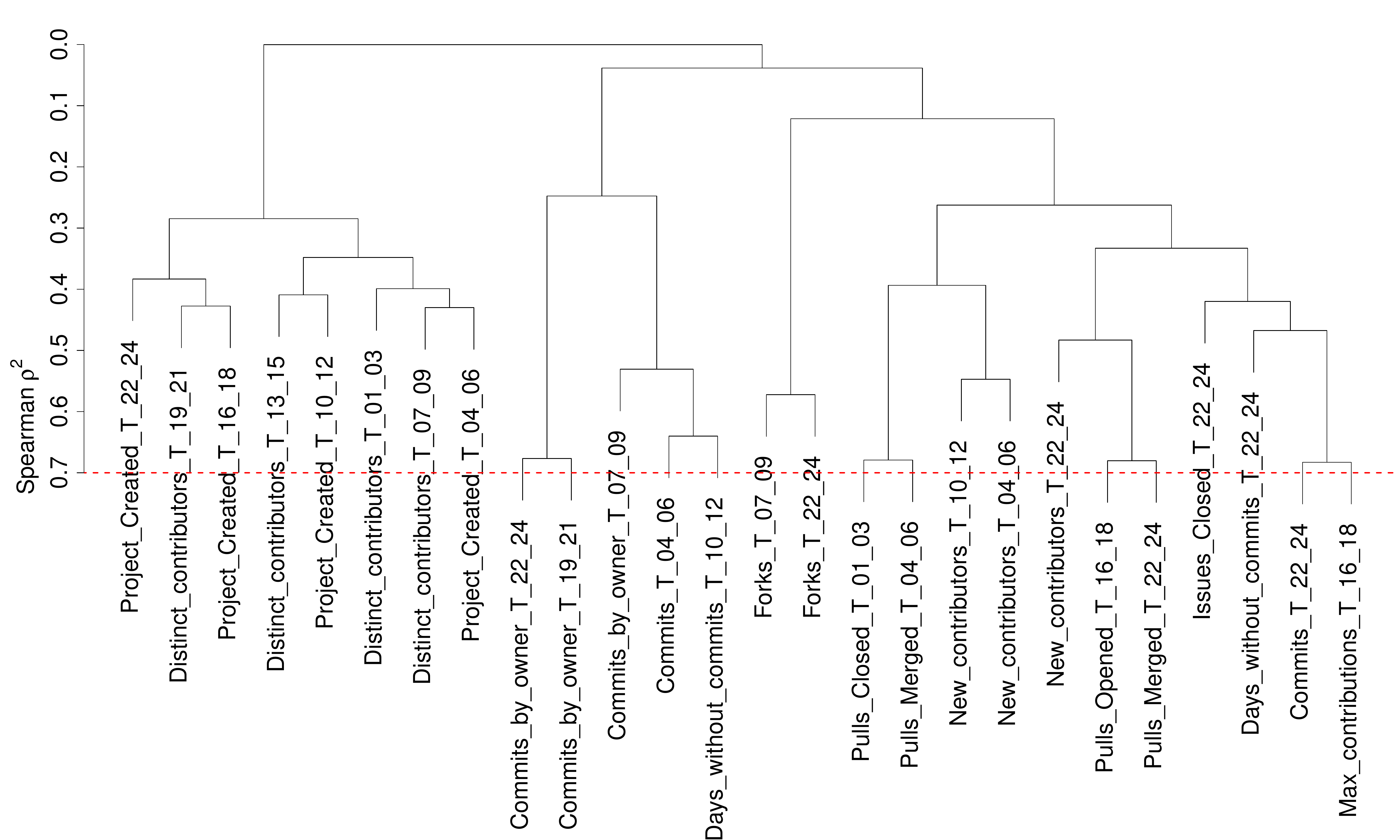}
\caption{Correlation analysis for the 104 data points collected for the features in scenario 8 (24 months, 3-month interval). 78 data points (75\%) are removed in this case, due to correlations with other data points, and therefore they do no appear in this final clustering.}
\label{fig:dendogram}
\end{figure}

\noindent{\bf Correlation Analysis.} As usual in machine learning experiments, we remove correlated features, following the process described by \citet{bao2017will}. To this purpose, we use a clustering analysis---as implemented in a R package named {\em Hmisc}\footnote{http://cran.r-project.org/web/packages/Hmisc/index.html}---to derive a hierarchical clustering of correlations among data points (extracted for the features in each scenario). For sub-hierarchies  with correlations larger than 0.7, we select only one data point for inclusion in the respective machine learning model, as common in other works~\citep{bao2017will, tian2015characteristics}.  
For example, Figure~\ref{fig:dendogram} shows the final hierarchical clustering for the scenario with 24 months, considering a 3-month interval (scenario~8). The  analysis in this scenario checks correlations among 104 data points (13 features $\times$ 8 data points per feature). As a result, 78 data points are removed due to correlations with other points and therefore do not appear in the dendogram presented in Figure~\ref{fig:dendogram}. Finally, Table~\ref{tab:features-removed} shows the total number and percentage of data points removed in each scenario, after correlation analysis. As we can see, the percentage of removed  points is relevant, ranging from 43\% (scenario 7) to 75\% (scenario 8).\\[-.2cm]
\begin{table}[!ht]
    \centering
    \caption{Total number and percentage of data points removed in each scenario, after correlation analysis.}    
    \begin{tabular}{c r r r r r r r r r r}
     \toprule
	\textbf{Scenario} & \textbf{1} & \textbf{2} & \textbf{3} & \textbf{4} & \textbf{5} & \textbf{6} & \textbf{7} & \textbf{8} & \textbf{9} & \textbf{10} \\
	\hline
    \# & 17 & 6  & 38 & 18 & 7  & 56 & 17 & 78 & 34 & 19   \\ 
    \% & 65 & 46 & 73 & 69 & 54 & 72 & 43 & 75 & 65 & 73   \\
    %\hline
        \bottomrule
    \end{tabular}
    \label{tab:features-removed}
\end{table}

\noindent{\bf Machine Learning Classifier.} 
We use the data points extracted in each scenario to train and test models for predicting whether a project is unmaintained. In other words, we train and test ten machine learning models, one for each scenario. After that, we select the best model/scenario to continue with the paper. Particularly, we use the Random Forest algorithm~\citep{breiman2001random} to train the models because it has several advantages, such as robustness to noise and outliers~\citep{tian2015characteristics}. In addition, it is adopted in many other software engineering works~\citep{menzies2013local, peters2013better,  provost2001robust,fse2016-andre}. We compare the result of Random Forest with two baselines: baseline \#1 (all projects are predicted as unmaintained) and baseline \#2 (random predictions). We use the Random Forest implementation provided by {\em randomForest}'s R package\footnote{https://cran.r-project.org/web/packages/randomForest/} and 5-fold stratified cross validation to evaluate the models effectiveness. In 5-fold cross validation, we randomly divide the dataset into five folds, where four folds are used to train a classifier and the remaining fold is used to test its performance. Specifically, stratified cross validation is a variant, where each fold has approximately the same proportion of each class~\citep{breiman2001random}. We perform 100 rounds of experiments and report average results.\\[-.2cm]

\noindent{\bf{Evaluation Metrics.}}  
When evaluating the projects in the test fold, each project has four possible outcomes: (1) it is truly classified as unmaintained (True Positive); (2) it is classified as unmaintained but it is actually an active project (False Positive); (3) it is classified as an active project but it is actually an unmaintained one (False Negative); and (4) it is truly classified as an active project (True Negative). Considering these possible outcomes, we use six metrics to evaluate the performance of a classifier: precision, recall, F-measure, accuracy, AUC (Area Under Curve), and Kappa, which are commonly adopted in machine learning studies~\citep{tian2015characteristics, tian2015automated, da2014empirical, lamkanfi2010predicting, lessmann2008benchmarking}. Precision and recall measure the correctness and completeness of the classifier, respectively. F-measure is the harmonic mean of precision and recall. Accuracy measures how many projects are classified correctly over the total number of projects. AUC refers to the area under the Receiver Operating Characteristic (ROC) curve. Finally, kappa evaluates the relationship between the observed accuracy and the expected one~\cite{bookML_R}, which is particularly relevant in imbalanced datasets, as the dataset used to train the machine learning models (\totalActiveProjects\ active projects {\em vs} \totalDeprecatedProjects\ unmaintained ones).

\begin{table*}[!t]
    \centering
     \small
    \caption{Prediction results (mean  of 100 iterations, using 5-cross validation); best results are in bold.}  
    \begin{tabular}{ l | r r | r r r | r r | r r r }
   
        \toprule
        
        \multirow{2}{*}{\bf Metrics} 
        & \multicolumn{2}{c | }{{\bf 0.5 Year}}
        & \multicolumn{3}{c | }{{\bf 1 Year}}
        & \multicolumn{2}{c | }{{\bf 1.5 Years}}
        & \multicolumn{3}{c  }{{\bf 2 Years}} \\
    	\multicolumn{1}{c | }{}
        & \multicolumn{1}{c|}{\textbf{3 mos}}	
        & \multicolumn{1}{c|}{\bf 6 mos}
        & \multicolumn{1}{c|}{\bf 3 mos}
        & \multicolumn{1}{c|}{\bf 6 mos}
        & \multicolumn{1}{c|}{\bf 12 mos}         
        & \multicolumn{1}{c|}{\bf 3 mos}
        & \multicolumn{1}{c|}{\bf 6 mos} 
        & \multicolumn{1}{c|}{\bf 3 mos}    
        & \multicolumn{1}{c|}{\bf 6 mos}
        & \multicolumn{1}{c}{\bf 12 mos} 
        \\
        \midrule	%		0.5/3		0.5/6	  1/3	 1/6	1/12  1.5/3	  1.5/6	  2/3	2/6  2/12
        Accuracy	& {0.90} & {0.91} & {0.91} & {0.90} & {0.89} & {0.91} &  {0.90} & {\bf 0.92} &  {0.91} & {0.90} \\
        Precision	& {0.83} & {0.87} & {0.87} & {0.84} & {0.82} & {0.86} & {0.83} & {\bf 0.86} & {0.85} & {0.83} \\
        Recall		& {0.78} & {0.74} & {0.77} & {0.75} & {0.72} & {0.78} & {0.76} & {\bf 0.81} & {0.79} & {0.73} \\
        F-measure 	& {0.80} & {0.79} & {0.81} & {0.79} & {0.77} & {0.82} & {0.79} & {\bf 0.83} & {0.82} & {0.78} \\
        Kappa		& {0.74} & {0.74} & {0.76} & {0.73} & {0.70} & {0.76} & {0.73} & {\bf 0.78} & {0.76} & {0.71} \\
        AUC			& {0.86} & {0.85} & {0.86} & {0.85} & {0.83} & {0.87} & {0.85} & {\bf 0.88} & {0.87} & {0.84} \\
        \bottomrule
    \end{tabular}
    \label{tab:prediction-results-scenarios}
\end{table*}

\subsection{Experimental Results} 
\label{subsec:experimental-results}
Table~\ref{tab:prediction-results-scenarios} shows the results for each scenario. As we can see, Random Forest has the best results (in bold) when the features are collected during 2 years, in intervals of 3 months.  In this scenario, precision is 86\% and recall is 81\%, leading to a F-measure of 83\%. Kappa is 0.78---usually, kappa values greater than 0.60 are considered quite representative~\cite{kappa}. Finally, AUC is 0.88, which is an excellent result in the Software Engineering domain~\citep{lessmann2008benchmarking, thung2012automatic, tian2015characteristics}. Table~\ref{tab:comparison-baselines} compares the results of the best scenario/model with baseline~\#1 (all projects are predicted as unmaintained) and baseline~\#2 (random predictions). Despite the baseline under comparison, there are major differences in all evaluation metrics. For example, F-measure is 0.37 (baseline \#1) and 0.30 (baseline \#2), against 0.83 (proposed model).

\begin{table}[!ht]
    \centering
    \caption{Comparison of the proposed machine learning model with  baseline \#1 (all projects are predicted as unmaintained) and baseline \#2 (random predictions).}    
    \begin{tabular}{ l r r r r}
        \toprule
       {\bf Metrics}    & \multicolumn{1}{c}{\bf Model} &  \multicolumn{1}{c}{\bf Baseline \#1}  &  \multicolumn{1}{c}{\bf Baseline \#2}\\ 
        \midrule
        Accuracy	   	& 	\sbar{92} {0.92}  	& \sbar{22} {0.22} 	& \sbar{49} {0.49} \\
        Precision	   	&  	\sbar{86} {0.86}	& \sbar{22} {0.22} 	& \sbar{22} {0.22} \\
        Recall	       	&  	\sbar{81} {0.81}	& \sbar{100} {1.00} & \sbar{48} {0.48}	\\
        F-measure 	   	&  	\sbar{83} {0.83}	& \sbar{37} {0.37} 	& \sbar{30} {0.30} \\
        Kappa		   	&	\sbar{78} {0.78}	& \sbar{0} {0.00} 	& \sbar{1} {0.01} \\     
        AUC		   		&	\sbar{88} {0.88}	& \sbar{50} {0.50} 	& \sbar{49} {0.49} \\   
        \bottomrule
    \end{tabular}
    \label{tab:comparison-baselines}
\end{table}

Random Forest produces a measure of the importance of the predictor features. Table~\ref{tab:features-importance} shows the top-5 most important features by Mean Decrease Accuracy (MDA), for the best model. Essentially, MDA measures the increase in prediction error (or reduction in prediction accuracy) after randomly shifting the feature values~\cite{calle2010letter,louppe2013understanding}.
As we can see, the most important feature is the number of commits in the last time interval (i.e., the interval delimited by months 22-24, $T_{22,24}$), followed by the maximal number of days without commits in the same interval and in the interval $T_{10,2}$.
As also presented in Table~\ref{tab:features-importance}, the first four features are related to commits; the first feature non-related with commits is the number of issues closed in the first time interval ($T_{1,3}$).
\begin{table}[!ht]
    \centering
    \caption{Top-5 most relevant features, by Mean Decrease Accuracy (MDA).}    
    \begin{tabular}{ l l r }
        \toprule
        {\bf Feature}					& {\bf Period}		& {\bf MDA} 	\\ 
        \midrule
        Commits							& T$_{22,24}$		& 	38.5	\\
        Max days without commits		& T$_{22,24}$		& 	28.6	\\
        Max days without commits		& T$_{10,12}$		& 	21.9	\\
        Max contributions by developer	& T$_{16,18}$		& 	21.1	\\
        Closed issues					& T$_{1,3}$			& 	18.0	\\
        \bottomrule
    \end{tabular}
    \label{tab:features-importance}
\end{table}

\section{Empirical Validation}
\label{sec:validation}

In this section, we {\em validate} the proposed machine learning model by means of a survey with the owners of projects classified as {\em unmaintained} and also with a set of deprecated GitHub projects.
Overall, our goal is to strengthen the confidence on the practical value of the model proposed in this work. Particularly, in this section we provide answers to first three research questions about this model:\\[-.3cm]

\noindent{\em RQ1: What is the precision according to GitHub developers?} \\[-.3cm]

\noindent{\em RQ2: What is the recall when identifying deprecated  projects?} \\[-.3cm]

\noindent{\em RQ3: How early does the  model identify unmaintained projects?} \\[-.3cm]

\subsection{Methodology}

\noindent{\bf RQ1:} To answer RQ1, we conduct a survey with GitHub developers. To select the participants, we first apply the proposed machine learning model in all projects from our dataset that were not used in the model's construction, totaling \testsetsize\ projects (\datasetsize\ $-$ \trainingsetsize\ projects). Then, we select \testsetClassifiedAsUnmaintained\ projects classified as unmaintained by the proposed model. From this sample, we remove 264 projects whose developers were recently contacted in our previous surveys~\cite{coelho2017why,coelho2018why}. We make this decision to not bother again these developers, with new e-mails and questions. Finally, we remove 2,270 projects whose owners do not have a public e-mail address on GitHub. As a result, we obtain a list of \totalSurveyParticipantsWithPublicEmail\ survey participants (2,856 $-$ 2,270 $-$ 264). However, before e-mailing these participants, the first author inspected the main page of each project on GitHub, to check whether it includes mentions to the project status, in terms of maintenance. We found \totalProjectsWithDeprecatedMessageInReadmeOfSurvey\ projects whose documentation states they are no longer maintained, by means of messages like this one:\\[-.3cm]

\noindent {\em This project is deprecated. It will not receive any future updates or bug fixes. If you are using it, please migrate to another solution.}\\[-.3cm]

Therefore, we do not send mails to the project owners, in such cases; and automatically consider these \totalProjectsWithDeprecatedMessageInReadmeOfSurvey\ projects as {\em unmaintained}.\\[-.2cm]

\noindent{\em Survey Period:} 
The survey was performed in the first two weeks of May, 2018. It is important to highlight that the machine learning model was constructed using data collected on November, 2017. Therefore, the {\em unmaintained} predictions evaluated in the survey refer to this date. We wait five months to ask the developers about the status of their projects because it usually takes some time until  developers actually accept the unmaintained condition of their projects. In other words, this section is based on predictions performed and available on November, 2017. However, these predictions are validated five months later, on May, 2018.\\[-.2cm]

\noindent{\em Survey Pilot and Questions:}  Initially, we perform a pilot survey with 75 projects ($\approx$ 25\%), randomly selected among the remaining 302 projects (\totalSurveyParticipantsWithPublicEmail\ $-$ \totalProjectsWithDeprecatedMessageInReadmeOfSurvey\ projects). We e-mail the principal developers of these projects with a single open question, asking them to confirm (or not) if their projects are unmaintained. We received 23 answers, which corresponds to a response ratio of 30.6\%. Then, two authors of this paper analyzed together the received answers to derive a list of recurrent themes. As a result, the following three common maintainability status were identified:\footnote{Project names are omitted, to preserve the respondent's anonymity; survey participants are identified by means of labels, in the format Pxx, where {\em xx} is an integer.}

\begin{enumerate}[leftmargin=*]

\item {\bf My project is under maintenance and new features are under implementation (6 answers):} As an example, we can mention the following answer:\\[-.3cm]

\noindent{\em [Project-Name] is still maintained. I maintain the infrastructure side of the project myself (e.g., make sure it's compatible with the latest Ruby version, coordinate PRs and issues, mailing list, etc.) while community provides features that are still missing. One such big feature is being developed as we speak and will be the highlight of the next release.} (P57)\\[-.3cm]

\item {\bf My project is finished and I only plan to fix important bugs (13 answers):} As an example, we mention the following answers:\\[-.3cm]

\noindent{\em It's just complete, at least for now. I still fix bugs on the rare occasion they are reported.} (P10)\\[-.3cm]

\noindent{\em I view it as basically ``done''. I don't think it needs any new features for the foreseeable future, and I don't see any changes as urgent unless someone discovers a major security vulnerability or something. I will probably continue to make changes to it a couple times per year, but mostly bug fixes.} (P68)\\[-.3cm]

\item {\bf My project is deprecated and I do not plan to implement new features or fix bugs (4 answers):} As an example, we can mention the following answer:\\[-.3cm]
 
\noindent{\em The project is unmaintained and I'll archive it.} (P74)

\end{enumerate}

After the pilot study, we proceed with the survey, by e-mailing the remaining 227 projects. However, instead of asking an open question---as in the pilot---we provide an objective question to the survey participants, about the maintainability status of their projects. In this objective question, we ask the participants to select one (out of three) status identified in the pilot study, plus an {\em other} option. This fourth option also includes a text form for the participants detailing their answers, if desired. Essentially, we change to an objective question format to make answering the survey easier, but without limiting the respondents' freedom to provide a different answer from the listed ones. In this final survey, we received 89 answers, representing a response ratio of 39.2\%. When considering both phases (pilot and final survey), we sent 302 e-mails, received \totalSurveyAnswersByDevelopers\ answers, representing an overall response ratio of 37.1\%. After adding the \totalProjectsWithDeprecatedMessageInReadmeOfSurvey\ projects that document their maintainability status, this empirical validation is based on \totalSurveyAnswersAndReadmeWithUnclearAnswers\ projects.\\[-.2cm]

\noindent{\bf RQ2:} To answer this second question, we construct a ground truth with projects that are no longer being maintained. First, we build a script to download the README (the main page of GitHub's projects) and search for a list of sentences that are commonly used to declare the unmaintained state of GitHub projects. This list is reused from our previous work~\cite{coelho2017why}, where we study the motivations for the failure of open source projects. It includes 32 sentences; in Table~\ref{tab:sentences}, we show a partial list, with 15 sentences.

\begin{table}[!ht]
    \centering
    \caption{Sentences documenting unmaintained projects.}  
 
    \begin{tabular}{l}
     \toprule
{\em no longer under development}, 
{\em no longer supported or updated},\\
    
{\em deprecation notice},
{\em dead project},
{\em deprecated},
{\em unmaintained},
\\
{\em no longer being actively maintained},
{\em not maintained anymore},
\\
{\em not under active development},
{\em no longer supported},
\\
{\em is not supported},
{\em is not more supported},
{\em no longer supported},
\\
{\em no new features should be expected},
{\em isn’t maintained anymore}\\
\bottomrule
    \end{tabular}
    \label{tab:sentences}
\end{table}

We searched (in May, 2018) for these sentences in the README of \testsetsize\ projects, which represent all \datasetsize\ projects selected for this work minus \trainingsetsize\ projects used in Section~\ref{sec:machine-learning}. In 451 READMEs (7.8\%) we found the mentioned sentences. Then, the first author of this paper carefully inspected each README, to confirm the sentences indeed refer to the project's status, in terms of maintenance. In the case of \totalProjectsUnmaintainedByReadme\ projects ($\approx$ 25\%), he confirmed this fact. Therefore, these projects are considered as a ground truth for this investigation. Usually, the unconfirmed cases refer to the deprecation of specific elements, e.g., methods or classes.\\[-.2cm]

\noindent{\bf RQ3:} To answer this third research question, we rely on projects whose unmaintained status, as predicted by the proposed model, is confirmed by the participants of the survey conducted in RQ1. Then, we compute  the number of days between November 30, 2018 (when the machine learning model proposed in this paper was built) and the last commit of the mentioned projects. For projects where this interval is less than one year, the proposed model is better than the strategy adopted in previous work~\cite{khondhu2013all,mens2014survivability, izquierdo2017empirical, coelho2017why}, which requires one year of commit inactivity to identify unmaintained projects.\\[-.2cm]

\begin{figure}[!t]
\centering
\includegraphics[width=10cm]{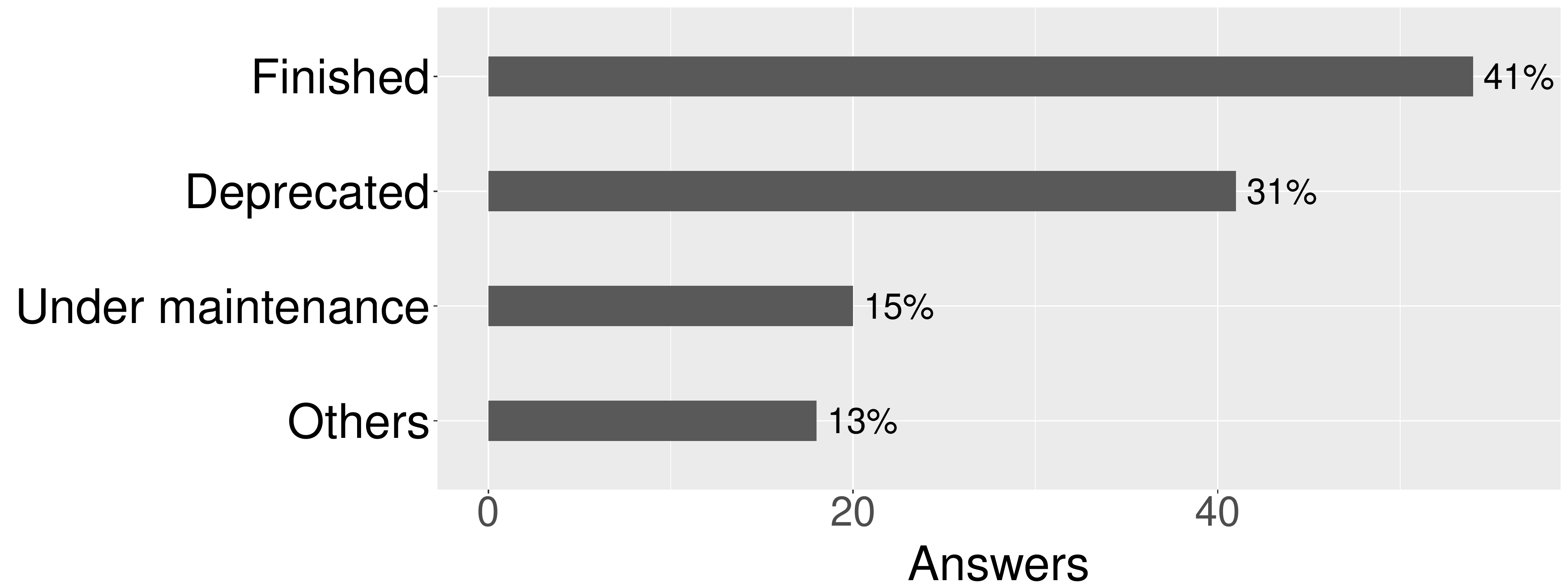}
\caption{Survey answers about projects' status.}
\label{fig:survey_answers}
\end{figure}

\subsection{Results}
\label{subsec:validation-results}

\noindent{\bf \large RQ1: Precision according to GitHub developers}\\[-.2cm]

Before presenting the precision results, Figure~\ref{fig:survey_answers} shows the survey results, including answers retrieved from the project's documentation, answers received in the pilot and answers received in the final survey. As we can see, the most common status, with \totalFinishedProjectsBySurveyAnswers\ answers (41\%) refers to {\em finished} projects, i.e., cases where maintainers see their projects as feature-completed and only plan to resume the maintenance work if relevant bugs are reported.\footnote{In our previous work~\cite{coelho2017why}, we also identified finished or completed open source projects. However, we argued these projects do not contradict Lehman's Laws about software evolution~\cite{lehman1980programs}, because they usually deal with a very stable or controlled environment (whereas Lehman's Laws focus on E-type systems, where {\em E} stands for evolutionary).} We also received \totalDeprecatedProjectsBySurveyAnswers\ answers (31\%) mentioning the projects are deprecated and no further maintenance is planned, including the implementation of new features and bug fixes. Finally, we received \totalOthersAnswersBySurvey\ answers in the {\em other} option. In this case, four participants did not describe their answer or provide unclear answers; furthermore, one participant mentioned his project is in a {\em limbo} state:\\[-.3cm]

\noindent{\em The status of [Project-Name] fits into a special category. Some of the tools it’s based on are either deprecated or not powerful enough for the goal of the project. This is part of the reason what’s keeping the project from being ``done''. I would call this status {\bf \em limbo}. (P24)}\\[-.3cm]

Seven participants answered their projects are {\em stalled}, as in this answer:\\[-.3cm]

\noindent{\em It is under going a rewrite... but has been {\bf \em stalled} based on my own priorities (P33)}\\[-.3cm]

To compute precision, we consider as {\em true positive} answers related to the following status: finished (\totalFinishedProjectsBySurveyAnswers\ answers), deprecated (\totalDeprecatedProjectsBySurveyAnswers\ answers), stalled (7 answers), and limbo (1 answer). The remaining answers are interpreted as {\em false positives}, including answers mentioning that new features are being implemented (\totalMaintainedProjectsBySurveyAnswers\ answers) and the answers associated to the fourth option ({\em other} option), but without including a description or with an unclear description (4 answers). By following this criteria, we received \totalTruePositivesAnswersBySurvey\ true positive answers and \totalFalsePositivesAnswersBySurvey\ false positive ones, resulting in a precision of \modelPrecision\%.\footnote{This computation of precision assumes that finished projects are unmaintained. However, we recognize that the risks of using finished projects might be lower than the ones faced when using deprecated or stalled projects.}

\begin{formal}
By validating the proposed model with \totalSurveyAnswersAndReadme\ GitHub developers, we achieve a {\bf precision} of \modelPrecision\%, which is a result very close to the one obtained when building the model (86\%). 
\end{formal}

\vspace*{0.2cm}
\noindent{\bf \large RQ2: Recall considering deprecated projects}\\[-.2cm]

The proposed machine learning model classifies 108 (out of \totalProjectsUnmaintainedByReadme) projects of the constructed ground truth as unmaintained, which represents a recall of \modelRecall\%. This value is significantly greater than the one computed when testing the model in Section~\ref{sec:machine-learning}. Probably, this difference is explained by the fact that only projects that are completely unmaintained expose this situation in their READMEs. Therefore, it is easier to detect and identify this condition.

\begin{formal}
By validating the proposed model with projects that declare themselves as unmaintained, we achieve a {\bf recall} of \modelRecall\%.
\end{formal}

\vspace*{0.2cm}
\noindent{\bf \large RQ3: How early can we detect unmaintained projects?}\\[-.2cm]

\begin{figure}[!t]
\centering
\includegraphics[width=7cm]{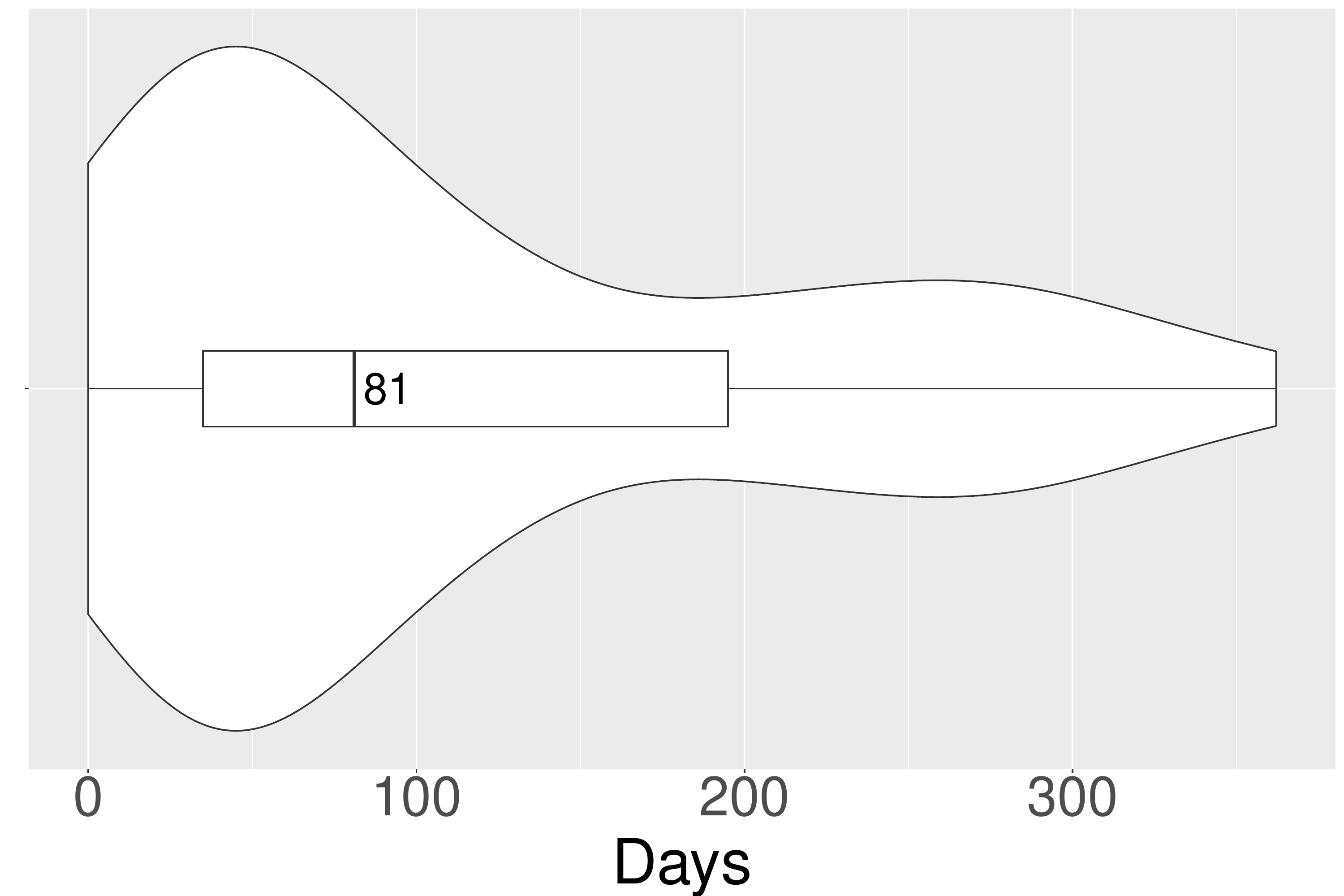}
\caption{Days since last commit for projects classified as unmaintained (considering the date of November, 2017, when the proposed model was computed).}
\label{fig:lma}
\end{figure}

\totalProjectsWithCommitsInTheLastYear\ (out of \totalTruePositivesAnswersBySurvey) projects classified as true positives by the surveyed developers have commits after November, 2016. Therefore, these projects would not be classified as unmaintained using the strategy followed in the literature, which requires one year of commit inactivity. In other words, in November, 2017, the proposed model classified \totalProjectsWithCommitsInTheLastYear\ projects as unmaintained, despite the existence of recent commits, with less than one year. Figure~\ref{fig:lma} shows a violin plot with the age of such commits, considering the date of November, 2017. The first, second, and third quartiles are 35, 81, and 195. Interestingly, for two projects the last commit occurred exactly on November, 30, 2018. Despite this fact, the proposed model classified these projects as unmaintained in the same date. If we relied on the standard threshold of one year without commits, these projects would have had to wait one year to receive this classification.

\begin{formal}
75\% of the studied projects are classified as unmaintained despite having recent commits, performed in the last year.
\end{formal}

\section{Characteristics of Unmaintained Projects}
\label{sec:characteristics-of-unmaintained}

In this section, we assess the characteristics of 2,856 projects classified as unmaintained by the proposed model. Although this model is not fully accurate, it showed a precision of 86\% in a dataset containing 754 active and 248 unmaintained projects (Section~\ref{subsec:experimental-results}).
Moreover, it achieved a precision of 80\%, when validated with the developers of 129 GitHub projects (Section~\ref{subsec:validation-results}). Therefore, this high precision measures---in different contexts---stimulated us to rely on the model to shed light on the characteristics of a large sample of unmaintained GitHub projects. 

Particularly, we provide answers to two research questions:\\[-.3cm]

\noindent{\em RQ4: How long does a GitHub project survive before become unmaintained?} The goal is to provide quantitative data about the lifetime of GitHub projects.\\[-.3cm]

\noindent{\em RQ5: How often unmaintained projects follow best OSS contribution practices?} With this question, the goal is to check whether common contribution practices contribute to extend the lifetime of GitHub projects.

\begin{table}[!t]
    \centering
    \caption{List of GitHub contribution practices.}
    \begin{tabular}{ p{4.5cm} p{10cm}}
        \toprule
        {\bf Maintenance Practice}			& {\bf Description}			\\ 
        \midrule
      License     & Presence of project's license (e.g.,~Apache, GNU, MIT)  \\\\
      Home Page   & Availability of a dedicated homepage in a non-GitHub-based domain   \\\\
      Continuous Integration    & Use of a continuous integration service (i.e.,~we check whether the projects use Travis CI, which is the most popular CI service on GitHub~\cite{hilton2016usage})    \\\\
      Contributing Guidelines   & Presence of contributing guidelines to help external developers make meaningful and useful contributions to the project  \\\\
      Issue Template    & Presence of an issue template (to instruct developers to write issues according to the repository's guidelines)  \\\\
      Code of Conduct   & Presence of a code of conduct, establishing expectations for behavior of the project's participants~\cite{tourani2017code}   \\\\
      Pull Request Template & Presence of a pull request template, which is a document to help developers to submit pull requests according to the repositories guidelines \\\\
      Support File  & Presence of a support file to direct contributors to specific support guidelines, such as community forums, FAQ documents, or support channels   \\\\
      First-timers-only issues  & Presence of labels recommending issues to newcomers (e.g.,~{\em help wanted} or {\em good-first-issue}) \\
        \bottomrule
    \end{tabular}
    \label{tab:github-pratices}
\end{table}

\subsection{Methodology}

\noindent{\bf RQ4:} To answer this fourth research question, we apply a survival analysis algorithm on \testsetClassifiedAsUnmaintained\ projects classified as unmaintained by our model. Survival analysis is a well-known technique, which is used for example on medical sciences to compute the probability of patients to stay alive for a certain number of days~\cite{cox2018analysis}. Moreover, survival analysis was successfully used in several software engineering studies~\cite{maldonado2017empirical},~\cite{valiev2018ecosystem},~\cite{samoladas2010survival},~\cite{lin2017developer}. In this work, we use survival analysis considering the lifetime of GitHub projects. In other words, we analyze the survival probability of a project over time. To determine the lifetime of a GitHub project, we compute the time difference between the first and last commit dates. Then, we use the Kaplan-Meier~\citep{kaplan1958nonparametric} non-parametric approximation to compute the survival curve, witch is the most widely used curve for estimating survival probabilities.\\[-.3cm]

\noindent{\bf RQ5:} To answer this last research question, we investigate whether projects classified as unmaintained follow (or not) a set of best open source contribution practices, which are recommended by GitHub.\footnote{\url{https://opensource.guide}} The rationale of this study is to compare the adoption of these practices between active and unmaintained projects. For each project, we collect the practices described in Table~\ref{tab:github-pratices}.\footnote{Seven of these contribution practices are explicitly recommended at: \url{
https://help.github.com/articles/helping-people-contribute-to-your-project}}

\subsection{Results}

\vspace*{0.2cm}
\noindent{\bf \large RQ4: How long does a GitHub project survive before become unmaintained?} \\[-.2cm]

\begin{figure*}[!h]
\centering
\includegraphics[width=9cm]{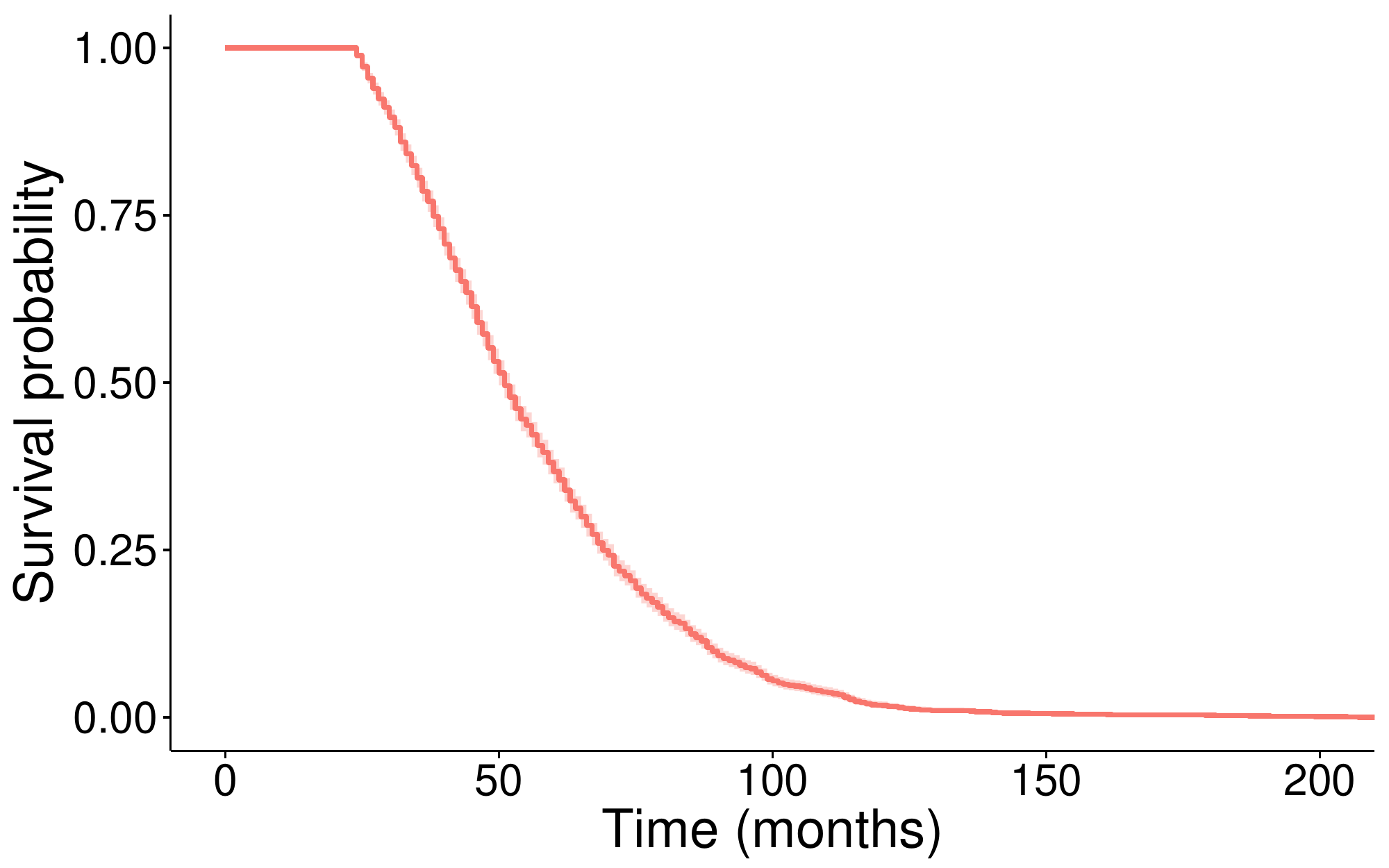}
\caption{Survival probability of unmaintained projects.}
\label{fig:survival-probability-unmaintained-by-model}
\end{figure*}

Figure~\ref{fig:survival-probability-unmaintained-by-model} shows the survival plot of \testsetClassifiedAsUnmaintained\ unmaintained projects, as classified by our model. As we only studied projects with at least 24 months, the survival curve is constant during this initial time frame. By inspecting Figure~\ref{fig:survival-probability-unmaintained-by-model}, we observe that the likelihood of an unmaintained project surviving for more than 50 months is close to 50\%. However, after 84 months (seven years) it declines to less than 10\%. In other words, 50\% of the unmaintained projects considered in this study moved to this state after 50 months ($\approx$ 4 years) and only 10\% remained active for more than 7 years.

Next, we investigate the characteristics of the projects with lower and higher survivability among the 2,856 unmaintained projects. The  projects  with lower survival probabilities are those in the first quartile, i.e., the lowest 25\% projects. In contrast,  the  projects  with higher survival probabilities are those in the fourth quartile, i.e., the highest 25\% projects.

\begin{figure*}[!t]
\centering
\subfloat[ref1][Contributors]
{
\includegraphics[width=5cm]{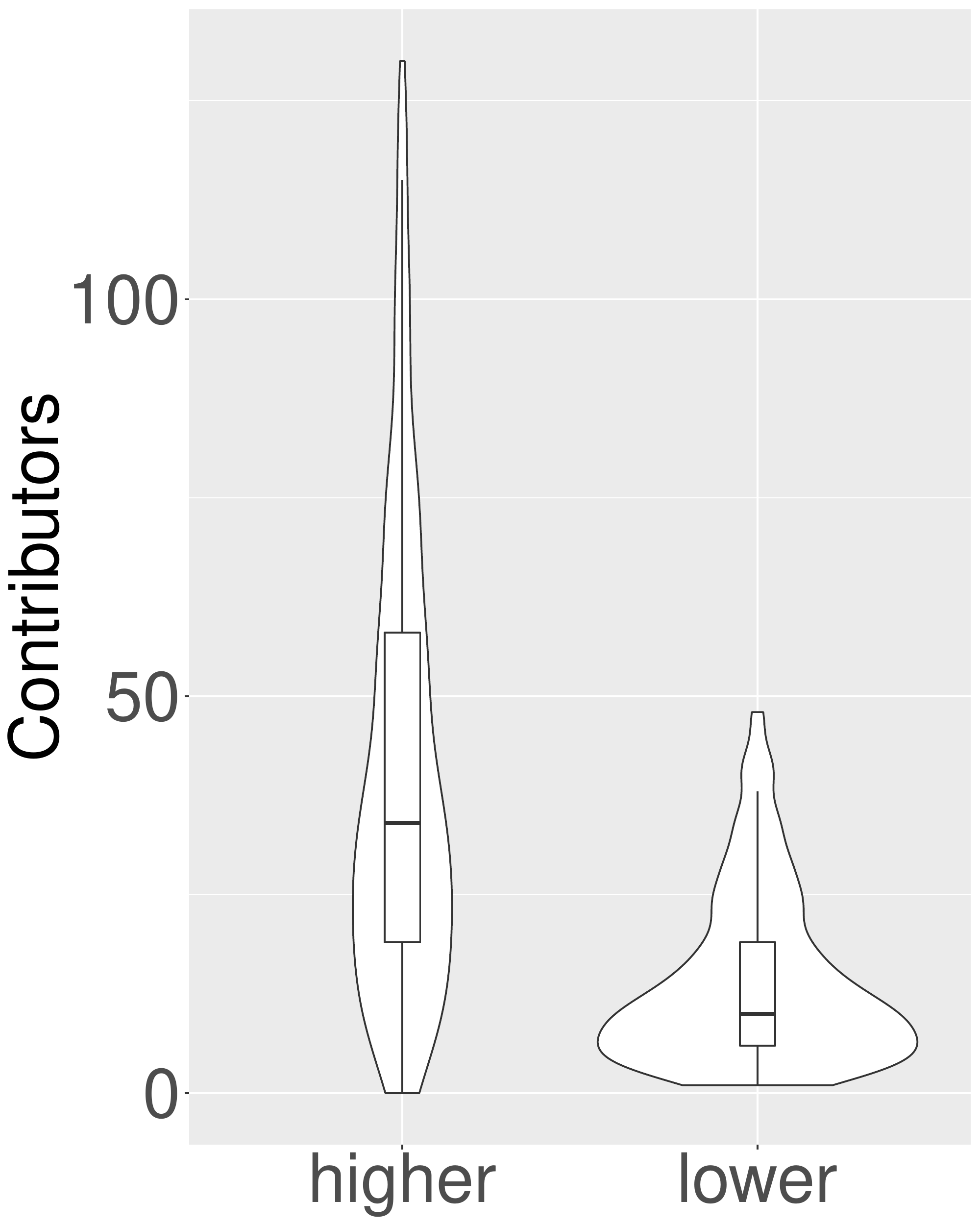}
}
\quad
\subfloat[ref2][Issues]
{
\includegraphics[width=5cm]{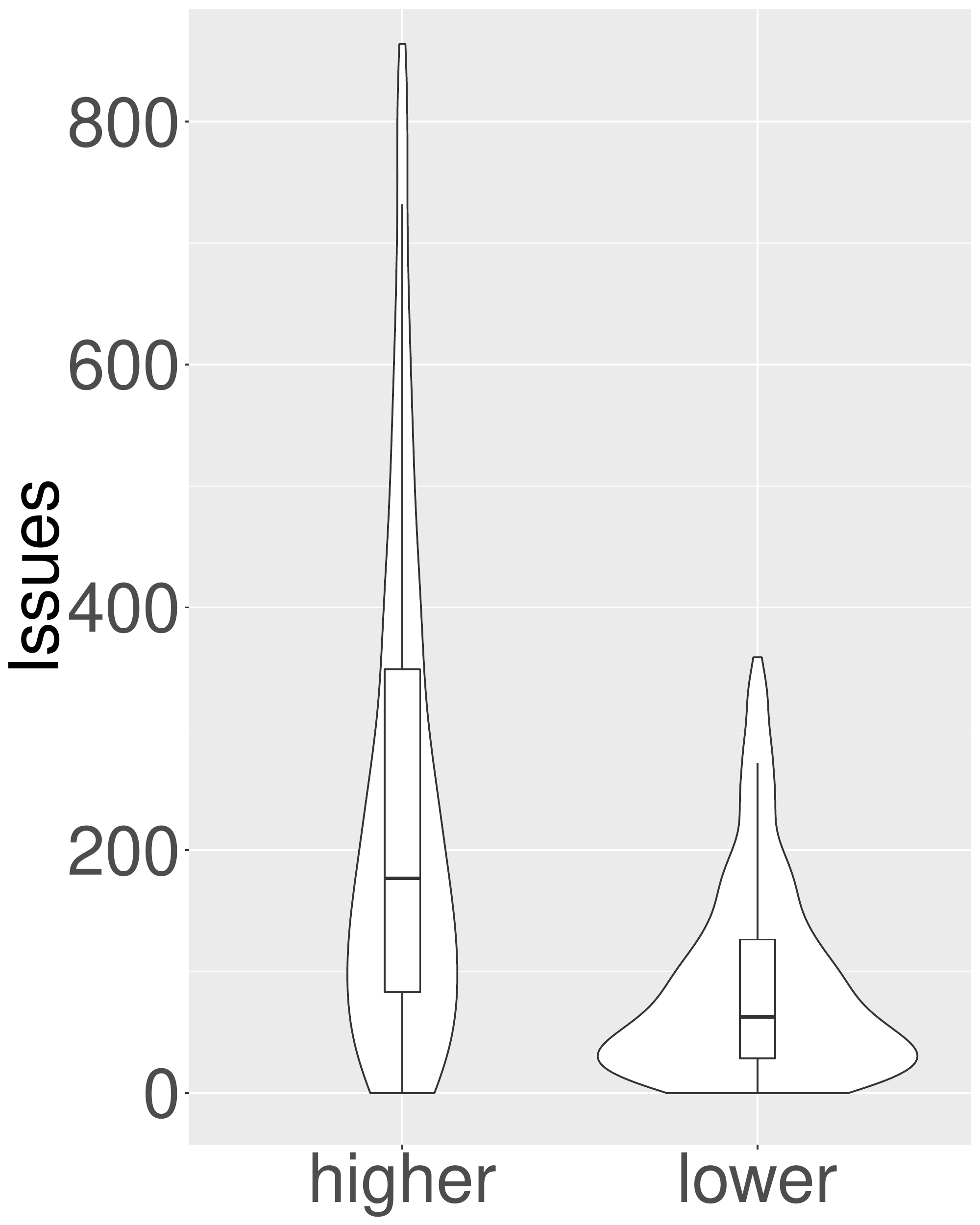}
}
\quad
\subfloat[ref3][Pulls]
{
\includegraphics[width=5cm]{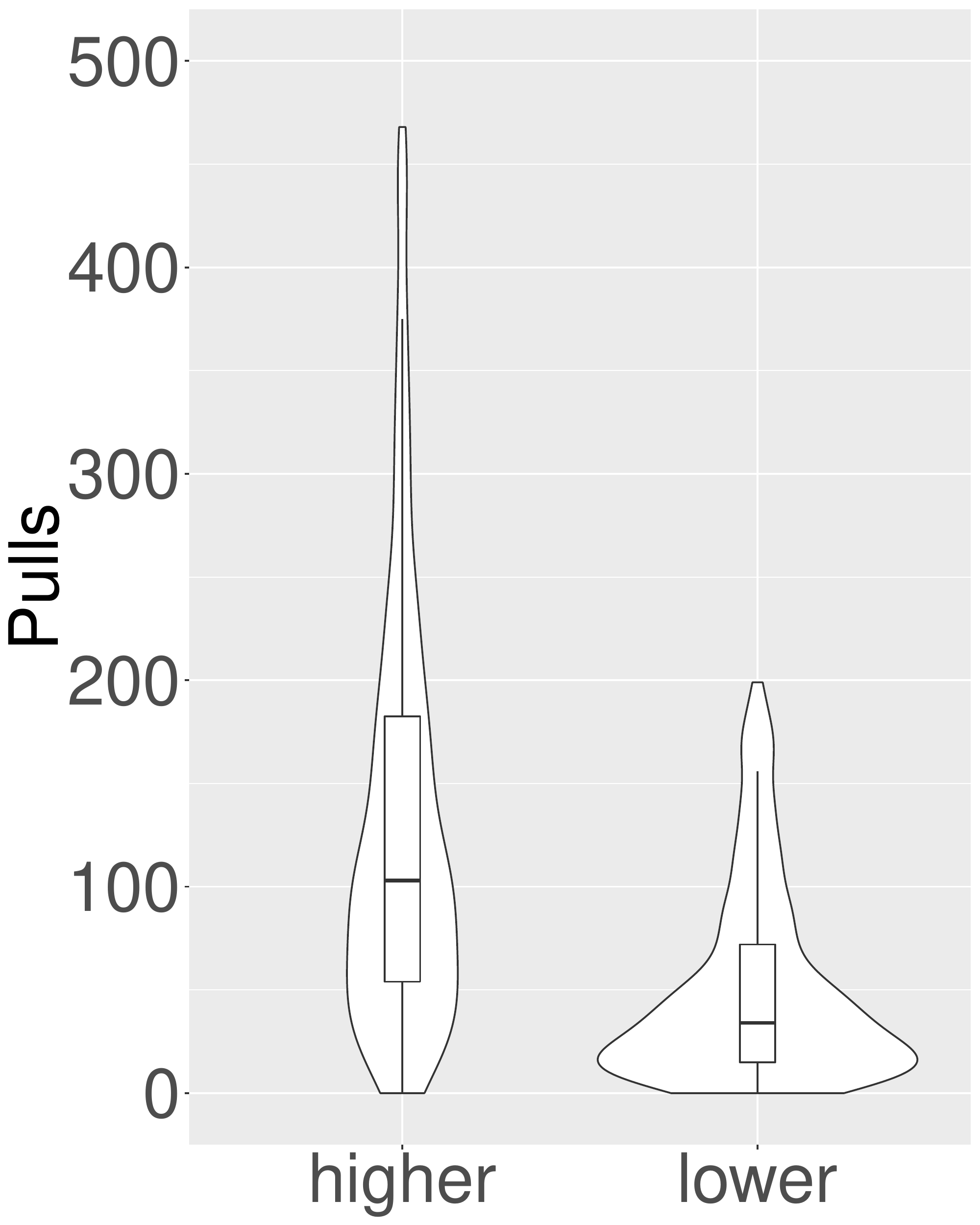}
}
\caption{Distribution of the projects by (a) contributors, (b) issues, and (c) pull requests.}
\label{fig:survival-lower-vs-higher}
\end{figure*}

Figure~\ref{fig:survival-lower-vs-higher} shows the distribution of the projects by (a) contributors, (b) issues, and (c) pull requests. As we can see, there is an important difference between the projects with higher and lower survivability on GitHub, in terms of total number of contributors, issues, and pull requests. These features seem to have an effect on the survivability of open source projects. Therefore, we reinforce the importance of always attract new contributors to GitHub projects. However, it is also important to consider that association does not imply causation. For example, by just attracting a high number of issues or pull requests, a project does not necessarily will have long survival.

Comparing the groups of projects with lower and higher survivability, we found a small effect size for continuous integration, followed by the adoption of contributing guidelines. For example, for projects with lower survivability, the percentage of projects following these practices are 38\% and 15\%, respectively. For the projects with higher survivability, the same values are 59\% and 23\%, respectively.

\begin{figure*}[!ht]
\centering
\subfloat[ref1][Account type]
{
\includegraphics[width=8cm]{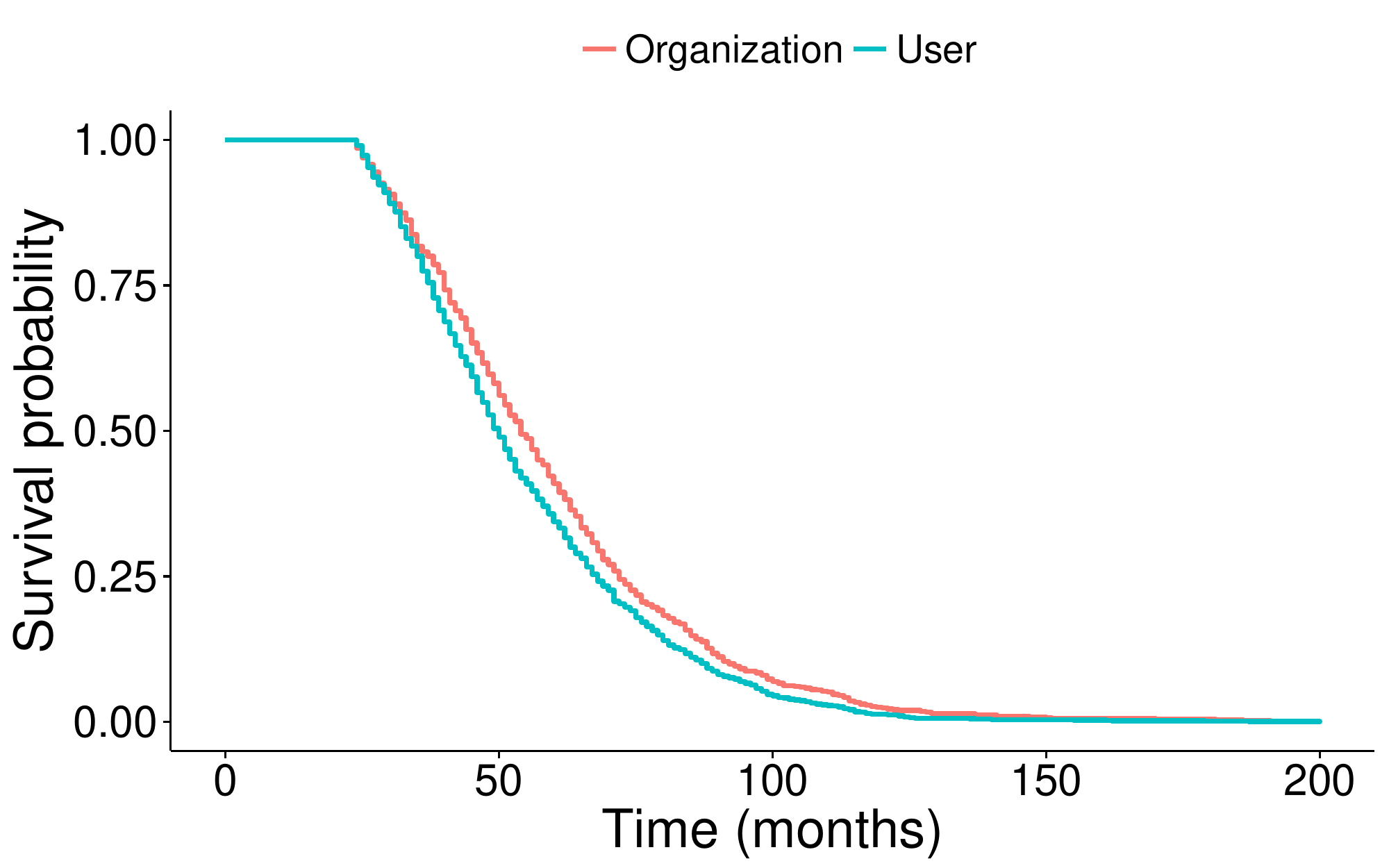}
}
\subfloat[ref2][Programming Language]
{
\includegraphics[width=8cm]{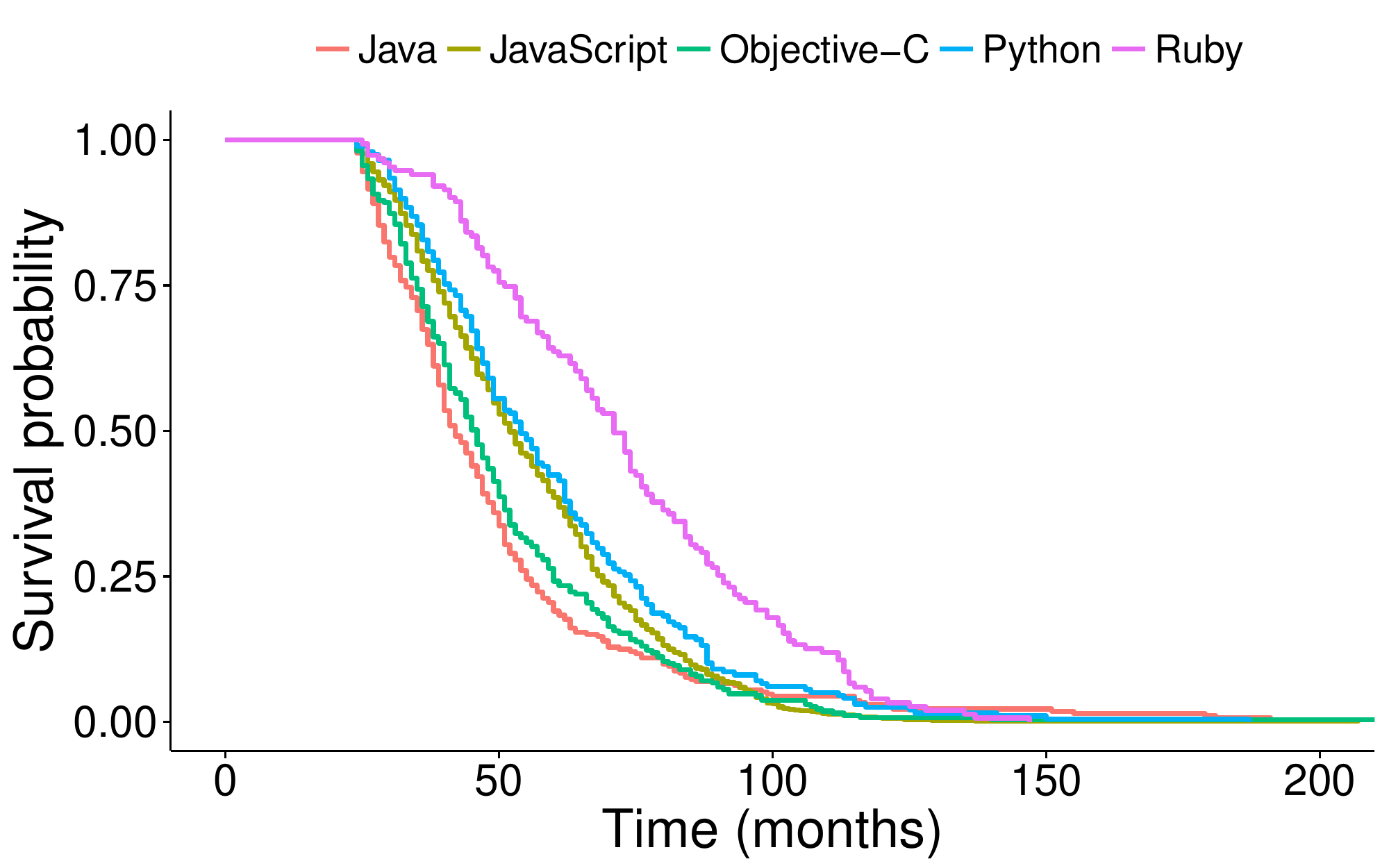}
}
\quad
\quad
\subfloat[ref3][Application Domain]
{
\includegraphics[width=8cm]{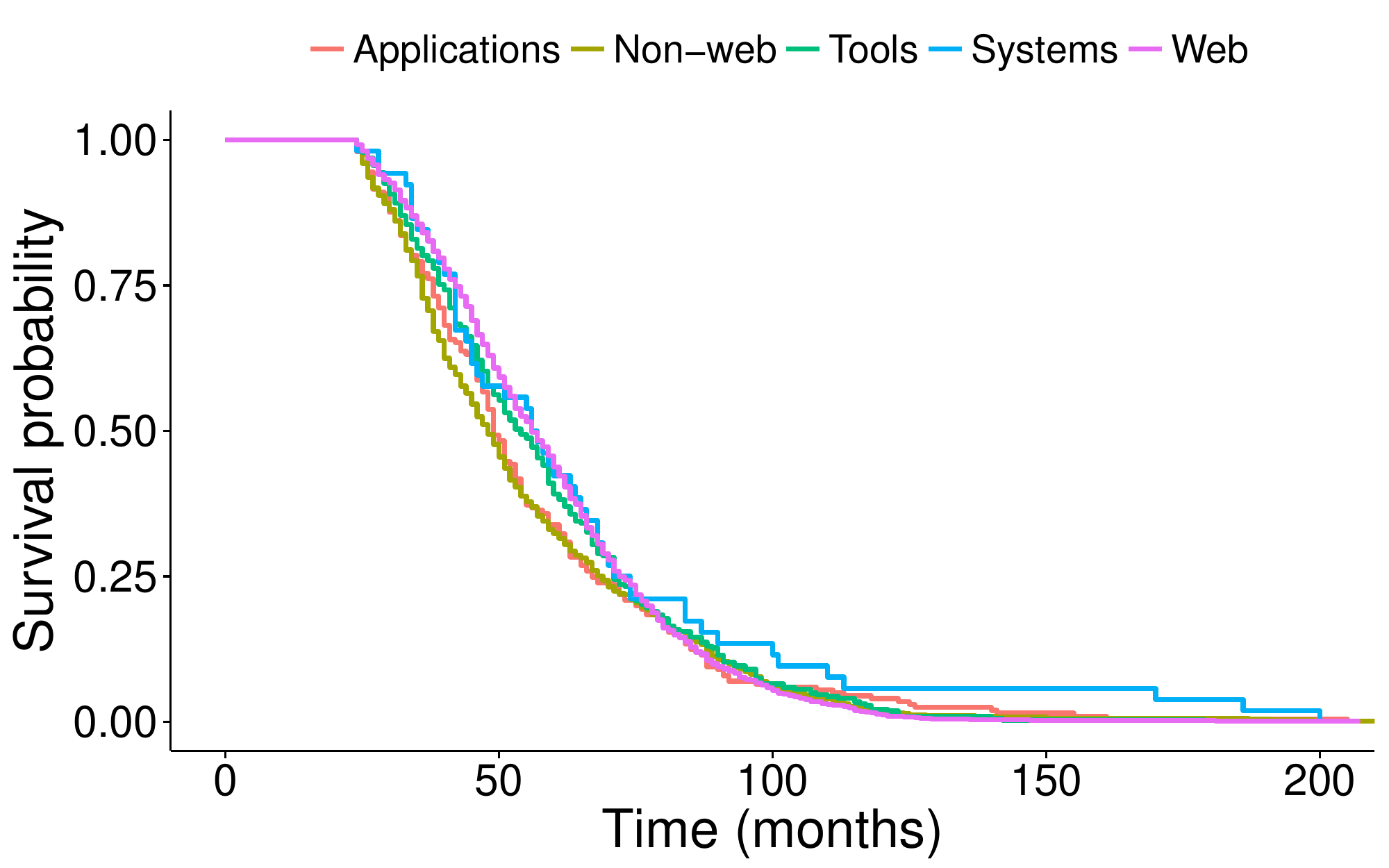}
}
\caption{Survival analysis by (a) GitHub account type, (b) programming language, and (c) application domain.}
\label{fig:survival-unmaintained-features}
\end{figure*}

We also generate specific survival plots for three projects features: account type, programming language, and application domain. Figure~\ref{fig:survival-unmaintained-features}a shows the survival plots for projects owned by organizations and individual users. As we can see, projects maintained by organizations have slightly greater survival probabilities than projects whose owner is an individual GitHub user. For example, after four years, the survival probabilities are 53\% and 60\%, for user and organization-owned projects, respectively. The distributions are statistically different, according to the one-tailed variant of the Mann-Whitney U test (p-value $\leq$ 0.05). However, we compute Cliff's delta (or $d$) to show the effect size of this difference and we found a negligible effect size. 

Figure~\ref{fig:survival-unmaintained-features}b shows the survival probabilities for the top-5 programming languages with more projects in our set of unmaintained projects. By applying Kruskal-Wallis test to compare multiple samples, we found that these distributions are different (p-value $\leq$ 0.05). The greatest difference happens between projects implemented in Java and Ruby. Particularly, projects implemented in Ruby tend to have higher survival probabilities than in other languages. By contrast, Java projects show lower survival probabilities. For example, after four years, the survival probabilities are 38\% and 79\%, for Java and Ruby projects, respectively.

Finally, Figure~\ref{fig:survival-unmaintained-features}c shows survival plots by application domain. To this propose, the first author of this paper manually classified the projects in five major application domains: {\em Application software}, {\em Non-web libraries and frameworks}, {\em Software tools}, {\em System software}, and {\em Web libraries and frameworks}. We reused these domains from a similar classification performed by~Borges et al.~\cite{borges2016icsme, borges2018stars}. The same domains are used in others studies~\cite{coelho2017why, borges2016promise}. By applying Kruskal-Wallis, the highest statistical difference occurs between {\em Non-web} and {\em Web} applications. For example, after four years, the survival probabilities are 50\% and 63\%, for {\em Non-web} and {\em Web} projects, respectively. Therefore, {\em Web} libraries tend to survive for more time than {\em Non-web} ones. Finally, we can also observe that {\em Systems software} have the highest survival probabilities. For example, after 8 years, the survival probabilities is twice than in other domains.

\begin{formal}
After characterizing the unmaintained projects, we found that there is a negligible difference on the survival probabilities of projects owned by individual and organizational accounts. Moreover, Ruby projects show higher probabilities of survival. Finally, {\em Systems Software} is the application domain with the highest survival probability.
\end{formal}

\noindent{\bf \large RQ5: How often unmaintained projects follow best OSS contribution practices?}\\[-0.3cm]

In this last RQ, we compare the adoption of a set of well-known contribution practices between active and unmaintained projects. First, we analyze the statistical significance of the difference in the usage of each practice between these groups of projects, by applying the Mann-Whitney test at p-value = 0.05. To show the effect size, we use Cliff's delta. Following the guidelines of previous work~\cite{grissom2005effect, tian2015mobile, linares2013api}, we interpret the effect size as small for $0.147 < d < 0.33$, medium for $0.33 \leq d < 0.474$, and large for $d \geq 0.474$.

Table~\ref{tab:unmaintained-vs-active-characteristics} shows the percentage of projects following each practice, considering {\em Active} vs {\em Unmaintained} projects. We found a {\em negligible} effect size for all practices, with the exception of continuous integration, contributing guidelines, and first-timers-only issues, when the effect size is {\em small}.

\begin{table}[!t]
\centering
\caption{Percentage of projects following recommended practices when maintaining GitHub repositories. The effect size reflects the extent of the difference between the unmaintained and active projects.}
\label{tab:unmaintained-vs-active-characteristics}
\begin{tabular}{ l  r  r r l }
\toprule
{\bf Maintenance Practice}   & \multicolumn{1}{c}{\bf Active} &  \multicolumn{1}{c}{\bf Unmaintained} & {\bf $d$} & {\bf Effect}	\\
\midrule
License                  	& \sbar{83} {0.83}  & \sbar{73} {0.73}	& 0.10  & negligible \\
Home Page                 	& \sbar{65} {0.65}  & \sbar{51} {0.51}	& 0.14  & negligible \\
Continuous Integration      & \sbar{71} {0.71}  & \sbar{45} {0.45}	& 0.26  & small \\
Contributing Guidelines     & \sbar{44} {0.44}  & \sbar{20} {0.20}	& 0.24  & small \\
Issue Template              & \sbar{8} {0.08}   & \sbar{2} {0.02}	& 0.06  & negligible \\
Code of Conduct             & \sbar{13} {0.13}  & \sbar{3} {0.03}	& 0.10  & negligible  \\
Pull Request Template       & \sbar{3} {0.03}   & \sbar{0} {0.00}   & 0.03  & negligible \\
Support File                & \sbar{1} {0.01}   & \sbar{1} {0.01} 	& 0.00  & negligible \\
First-timers-only issues    & \sbar{53} {0.53}  & \sbar{31} {0.31}  & 0.22  & small \\
\bottomrule
\end{tabular}
\end{table}

\begin{formal}
There is a small effect size between active and unmaintained projects for continuous integration, followed by the adoption of contributing guidelines, and the presence of labels recommending issues to newcomers. In contrast, the remaining practices revealed a negligible effect or does not exist in statistical terms.
\end{formal}

\section{Level of Maintenance Activity}
\label{sec:level-of-maintenance-activity}

In this section, we define a metric to express the {\em level of maintenance activity} of GitHub projects, i.e., a metric that reveals how often a project is being maintained. The goal is to alert users about projects that although classified as active by the proposed model are indeed close to an unmaintained status.

\subsection{Definition}
\label{sec:walking-dead-methodology}

The proposed machine learning model---generated by Random Forest---consists of multiple decision trees built randomly. Each tree in the ensemble determines a prediction to a target instance and the most voted class is considered as the final output. One possible prediction type of the Random Forest is the matrix of class probabilities. This matrix represents the proportion of the trees' votes. For example, projects predicted as {\em active} have probability $p$ ranging from 0.5 to 1.0. If  $p = 0.5$, the project is very similar to an unmaintained project; by contrast, $p = 1.0$ means the project is actively maintained. Using these probabilities, we define the {\em level of maintenance activity (LMA)} of a GitHub project as follows:
\[ LMA  =  2 * (p - 0.5) * 100 \]
This equation simply converts the probabilities $p$ computed by Random Forest to a range from 0 to 100; LMA equals to 0 means the project is very close to an unmaintained classification (since $p = 0.5$); and LMA equals to 100 denotes a project that is actively maintained (since $p = 1.0$). We do not calculate LMA for unmaintained projects, i. e., projects that received a random forest probability estimation lower than 0.5. 

\subsection{Results}
\label{sec:lma-results}
\begin{figure}[!ht]
\centering
\includegraphics[width=7cm]{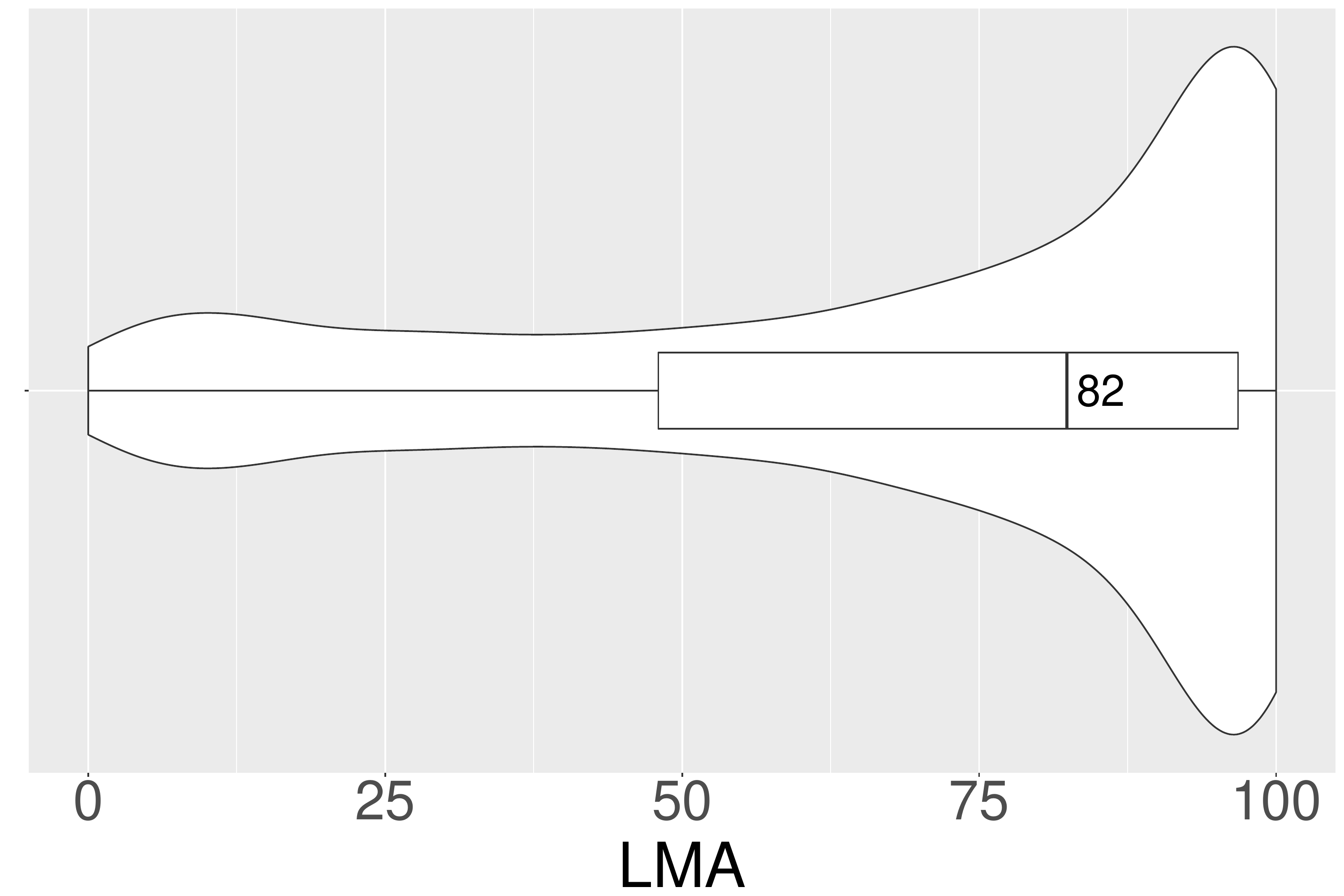}
\caption{Level of maintenance activity (LMA).}
\label{fig:active_projects_score}
\end{figure}

Figure~\ref{fig:active_projects_score} shows the LMA values for each project predicted as {\em active} (\testsetClassifiedAsActive\ projects, after excluding the projects used to train and test the proposed model, in Section~\ref{sec:machine-learning}). The first, second, and third quartiles are 48, 82, and 97, respectively. In other words, most studied projects are under constant maintenance (median 82). Indeed, 171  projects (5.8\%) have a maximal LMA, equal to 100. This list includes well-known and popular projects such as {\sc twbs/bootstrap},  {\sc meteor/meteor}, {\sc rails/rails}, {\sc webpack/web\-pack}, and {\sc elastic/e\-las\-ticsearch}.

\begin{figure*}[!t]
\centering

\subfloat[ref1][Commits]
{
\includegraphics[width=8cm]{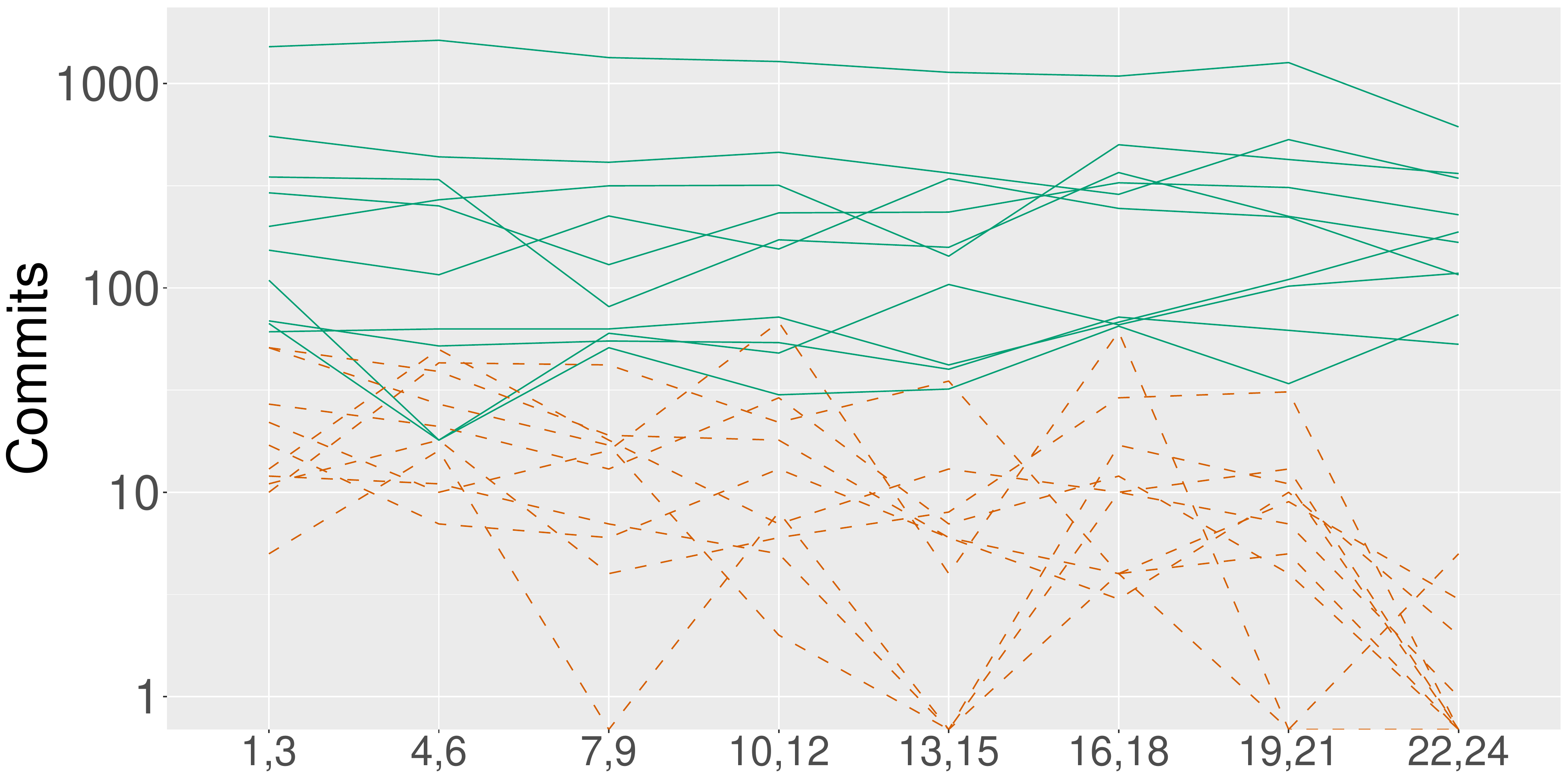}
}
\subfloat[ref2][Issues]
{
\includegraphics[width=8cm]{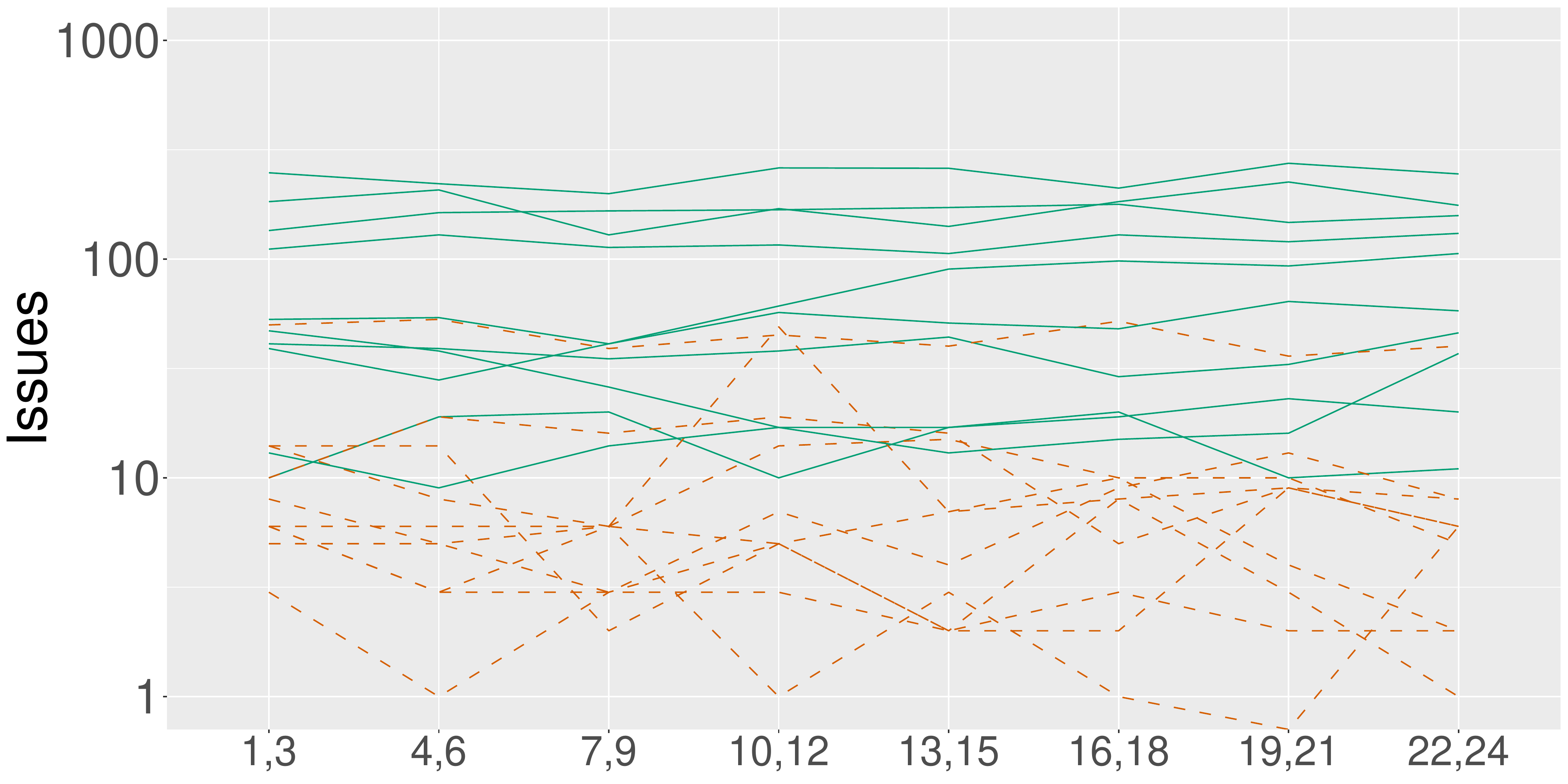}
}
\quad
\quad
\subfloat[ref3][Pull requests]
{
\includegraphics[width=8cm]{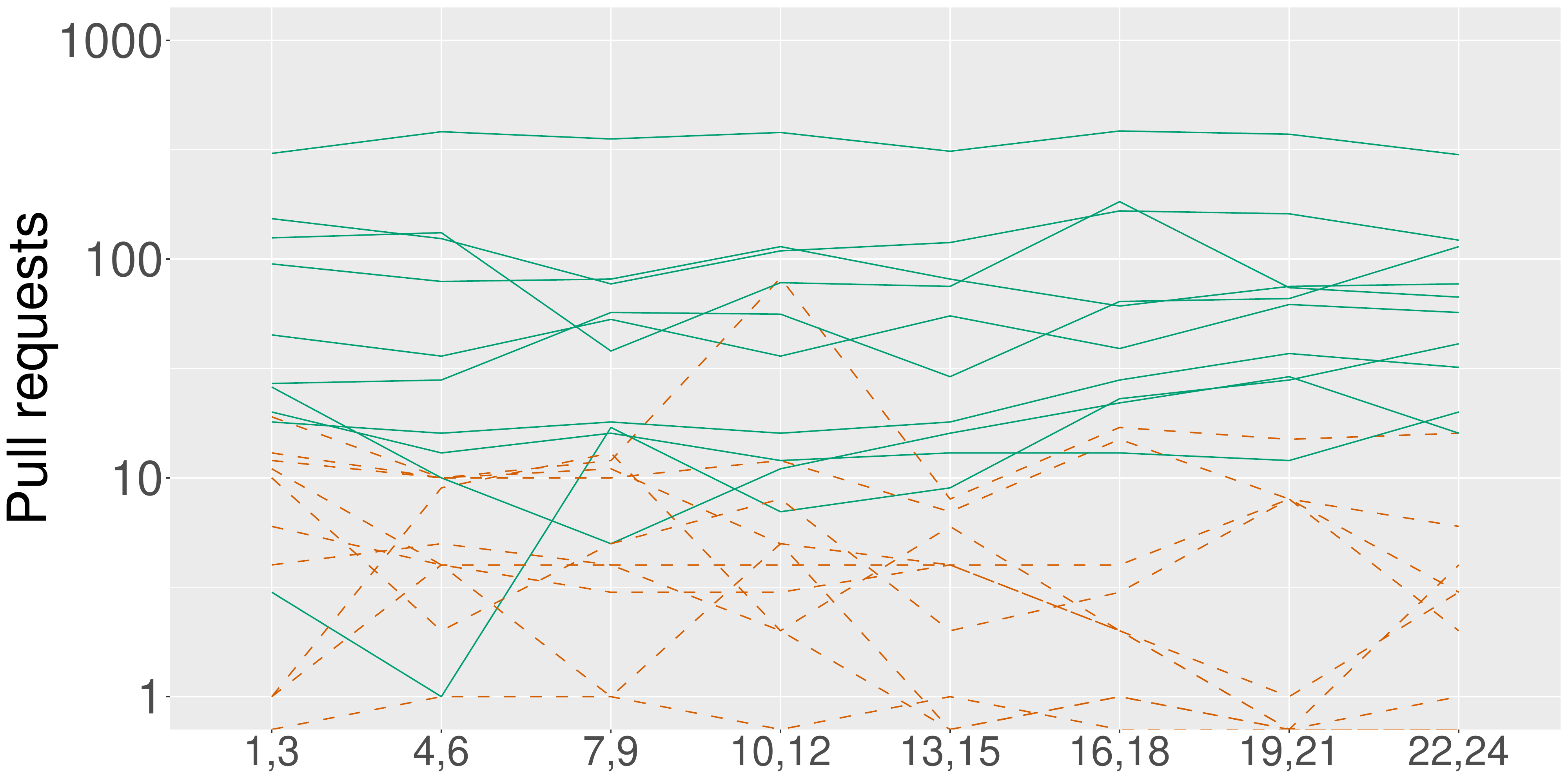}
}
\subfloat[ref4][Forks]
{
\includegraphics[width=8cm]{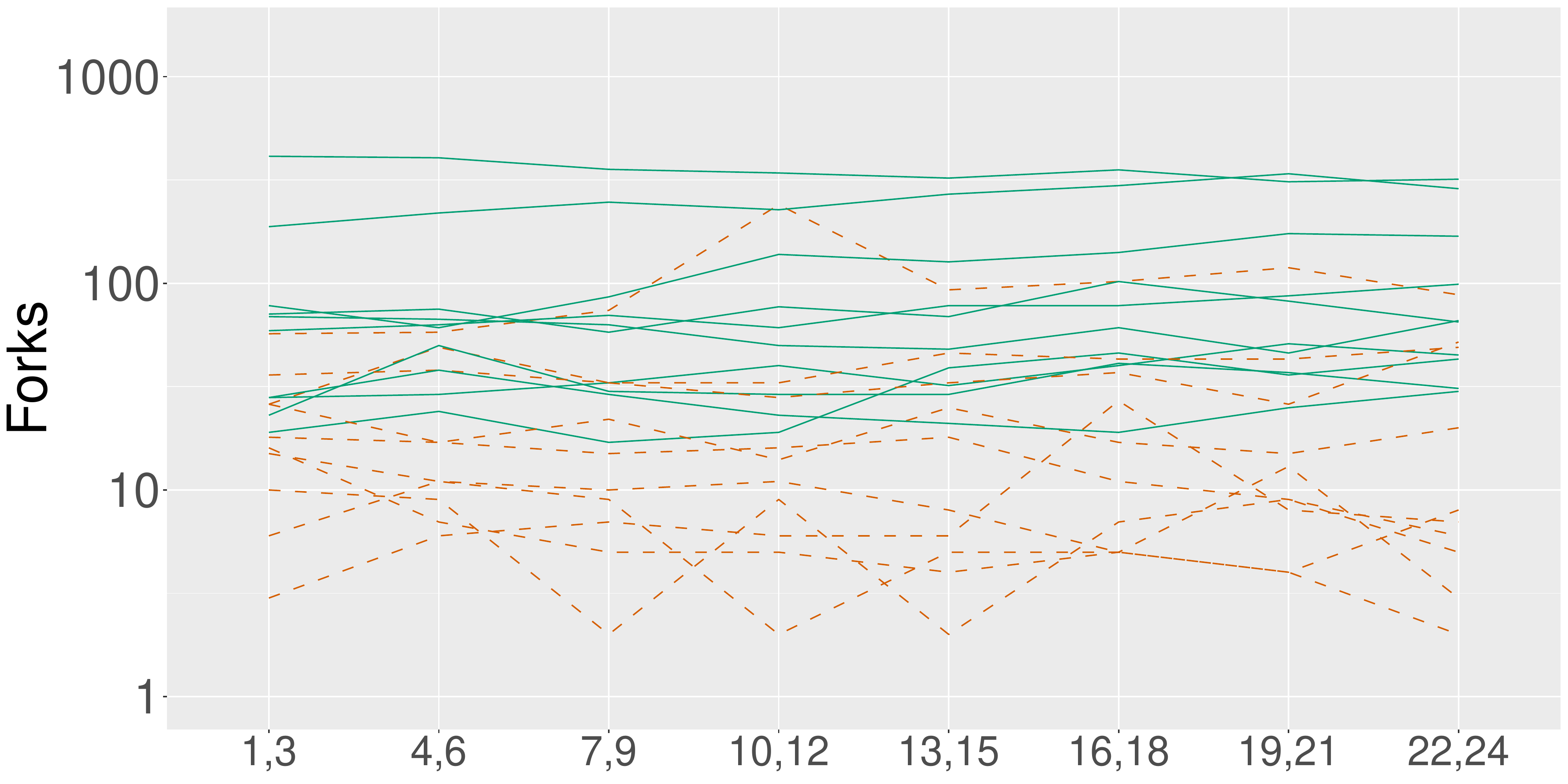}
}
\caption{Number of commits, issues, pull request, and forks over time of ten projects with maximal LMA (green lines) and ten projects with the lowest LMA in our dataset (red, dashed lines). Metrics are collected in intervals of 3 months (x-axis).}
\label{fig:active-vs-deprecated}
\end{figure*}

Figure~\ref{fig:active-vs-deprecated} compares a random sample of 10 projects with LMA equals to 100 (actively maintained, therefore) with ten projects with the lowest LMA (0 $\leq$ LMA $\leq$ 0.4). These projects are compared using  number of commits (Figure~\ref{fig:active-vs-deprecated}a), number of issues (Figure~\ref{fig:active-vs-deprecated}b), number of pull requests (Figure~\ref{fig:active-vs-deprecated}c), and number of forks (Figure~\ref{fig:active-vs-deprecated}d), in the last 24 months. Each line represents the project's metric values. The figures reveal major differences among the projects, regarding these metrics. Usually, the projects with high LMA present high values for the four  considered metrics (commits, issues, pull requests, and forks), when compared with projects with low LMA. In other words, the figures suggest that LMA plays an aggregator role of maintenance activity over time.

\begin{figure*}[!t]
\centering

\subfloat[ref1][LMA vs Stars ($\rho$ = 0.10)]
{
\includegraphics[width=8cm]{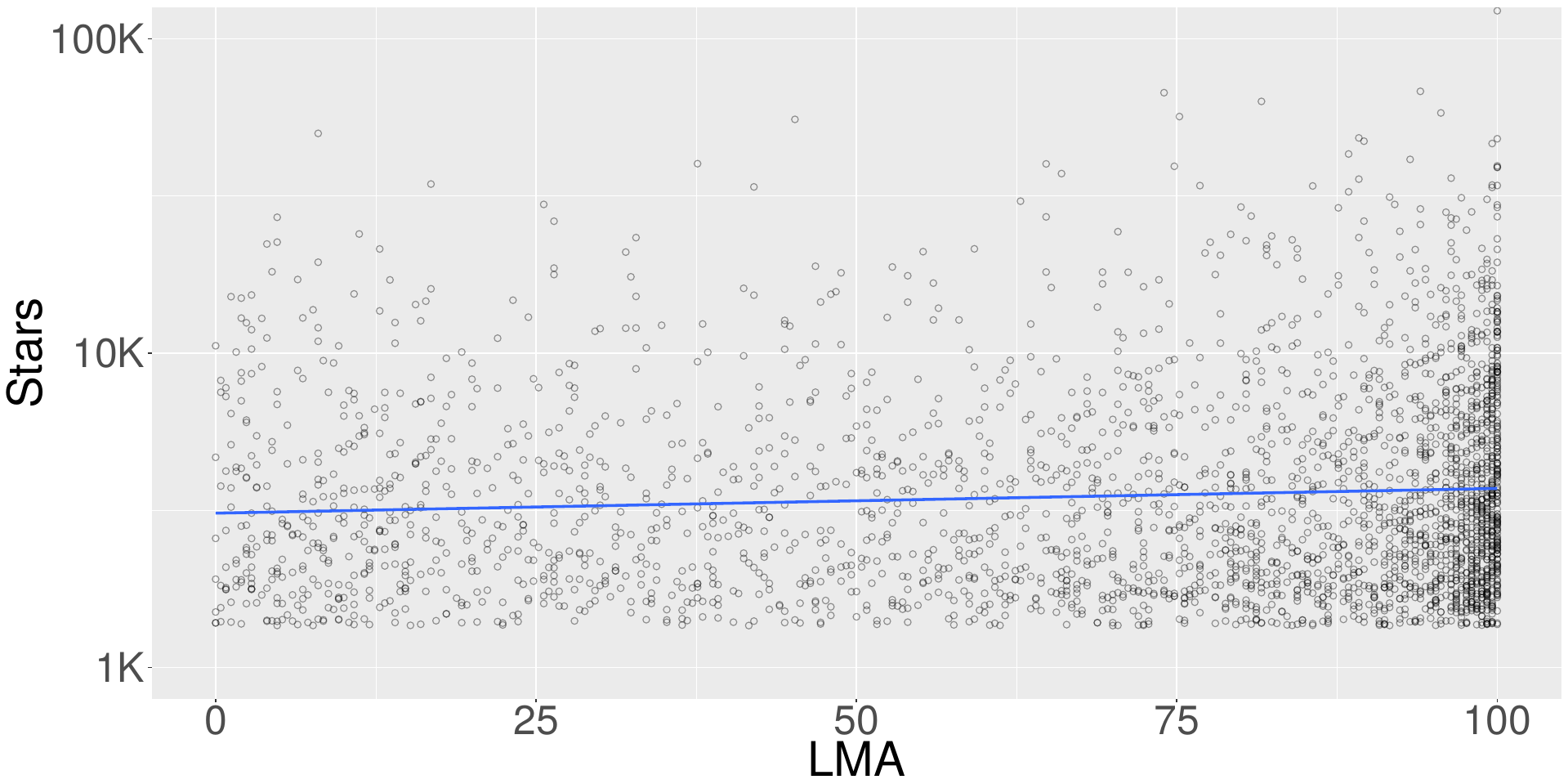}
}
\subfloat[ref2][LMA vs Contributors ($\rho$ = 0.44)]
{
\includegraphics[width=8cm]{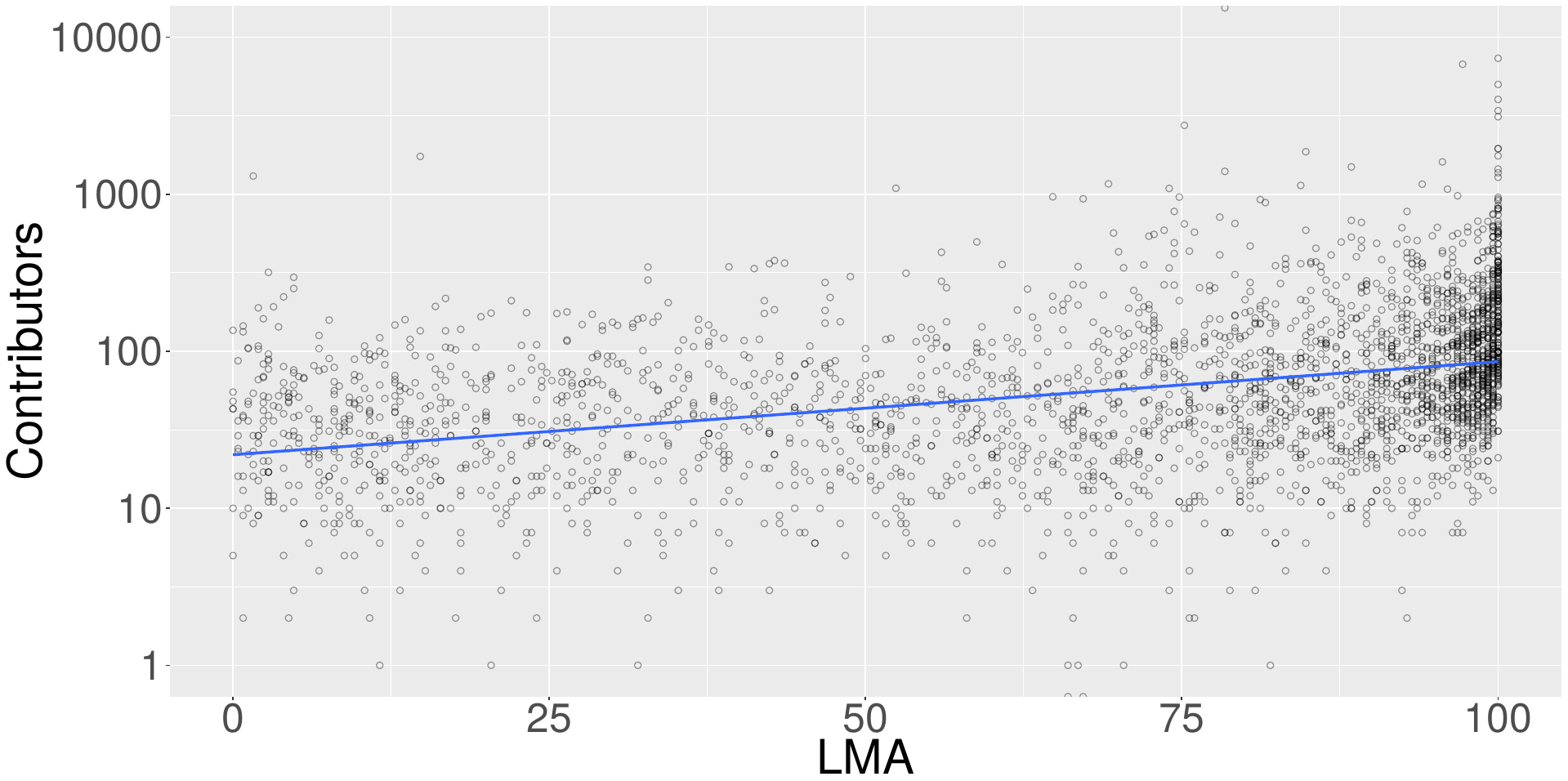}
}
\quad
\quad
\subfloat[ref3][LMA vs Core contributors ($\rho$ = 0.15)]
{
\includegraphics[width=8cm]{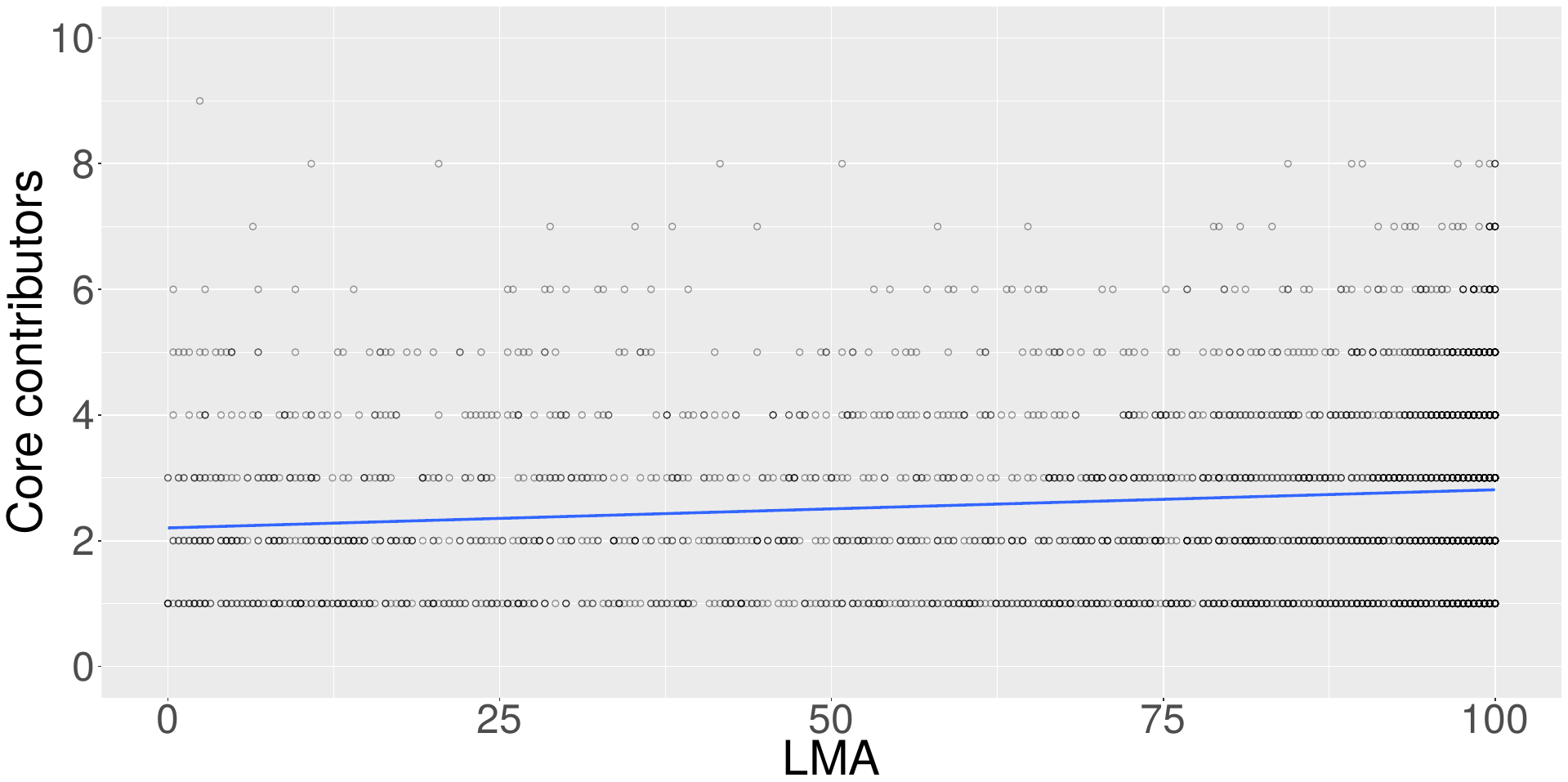}
}
\subfloat[ref4][LMA vs LOC ($\rho$ = 0.38)]
{
\includegraphics[width=8cm]{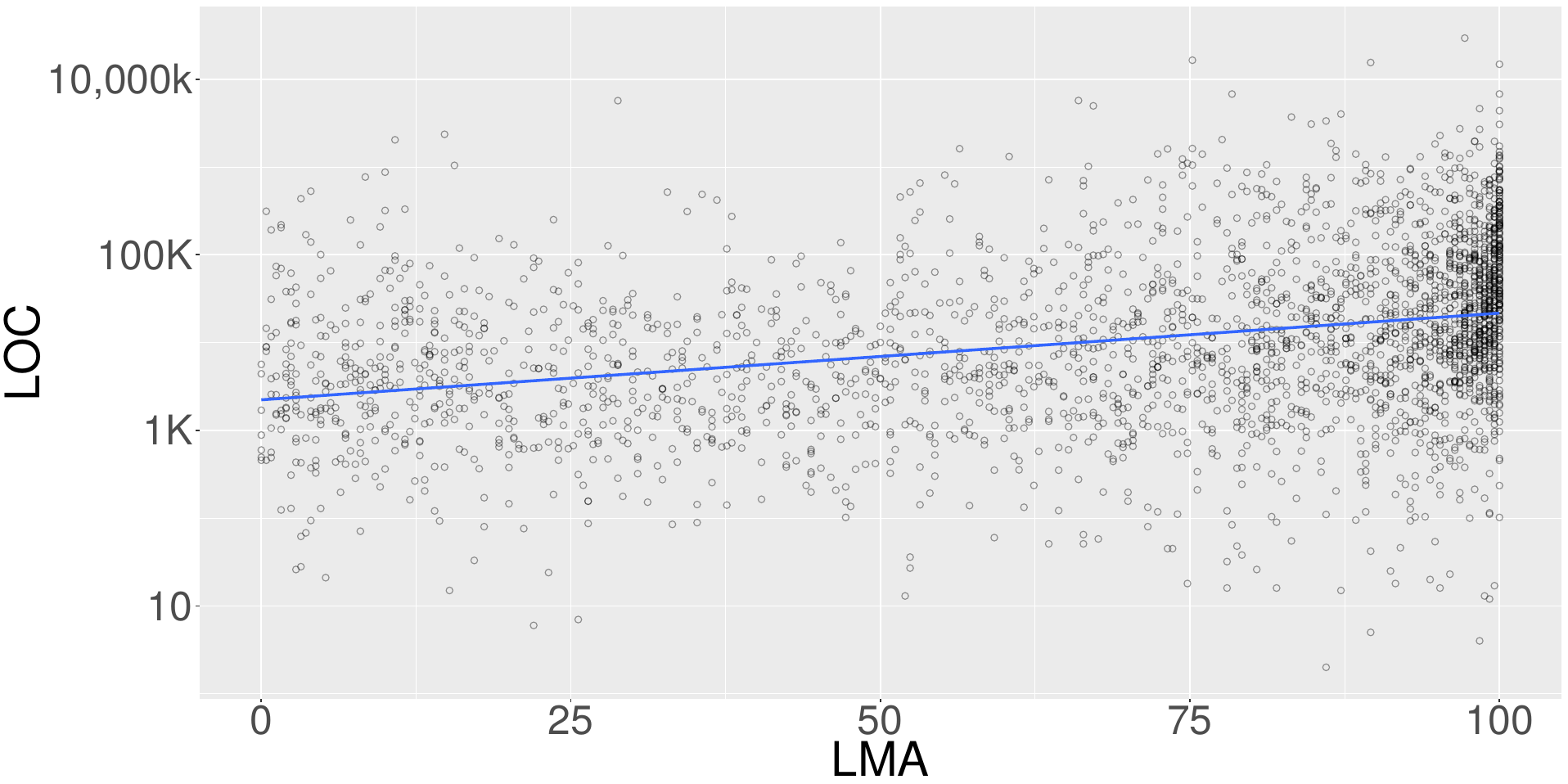}
}
\caption{Correlating LMA with (a) stars, (b) contributors, (c) core contributors, and (d) size. Spearman's $\rho$ is also presented.}
\label{fig:lma-vs-features}
\end{figure*}

Figure~\ref{fig:lma-vs-features} shows scatter plots correlating LMA and number of stars, contributors, core contributors, and size (in LOC) of projects classified as {\em active}. To identify core contributors, we use a common heuristic described in the literature: core contributors are the ones responsible together for at least 80\% of the commits in a project~\cite{koch2002effort,mockus2002two,robles2009evolution}. To measure the size of the projects, in lines of code, we used the tool {\sc AlDanial/cloc}\footnote{https://github.com/AlDanial/cloc}, considering only the programming languages in the TIOBE list.\footnote{https://www.tiobe.com/tiobe-index} We also compute Spearman's rank correlation test for each figure. The correlation of stars and core contributors is very weak ($\rho$ = 0.10 and $\rho$ = 0.15, respectively); with size, the correlation is weak ($\rho$ = 0.38); and with contributors, it is moderate ($\rho$ = 0.44); all p-values are less than 0.01. Therefore, it is common to have highly popular projects, by number of stars, presenting both low and high LMA values. For example, one project has 50,034 stars, but LMA = 8. A similar effect happens with size. For example, one project has $\approx$2 MLOC, but LMA = 10.8. The highest correlation is observed with contributors, i.e.,~projects with more contributors tend to have higher levels of maintenance activity.

\subsection{Validation with False Negative Projects}

In Section~\ref{sec:validation}, we found four projects that although declared by their developers as {\em unmaintained} are predicted by the proposed machine learning model as {\em active}. Therefore, these projects are considered false negatives, when computing recall. Two of such projects have a very low LMA:
{\sc nicklockwood/iRate} (LMA = 2) and {\sc gorangajic/react-icons}	(LMA = 12). Therefore, although predicted as {\em active},  these projects are similar to projects classified as {\em unmaintained}, as suggested by their low LMA.
A second project has an intermediate LMA value: {\sc spotify/HubFramework}	(LMA = 39.2). Finally, one project {\sc Homebrew/homebrew-php} has a high LMA value (LMA = 99.2). However, this project was migrated to another repository, when facing continuous maintenance.
In other words, in this case, the GitHub repository was deprecated, but not the project; therefore, {\sc Homebrew/homebrew-php} is a false, false negative (or a true negative).

\begin{figure}[!t]
\centering
\includegraphics[width=7cm]{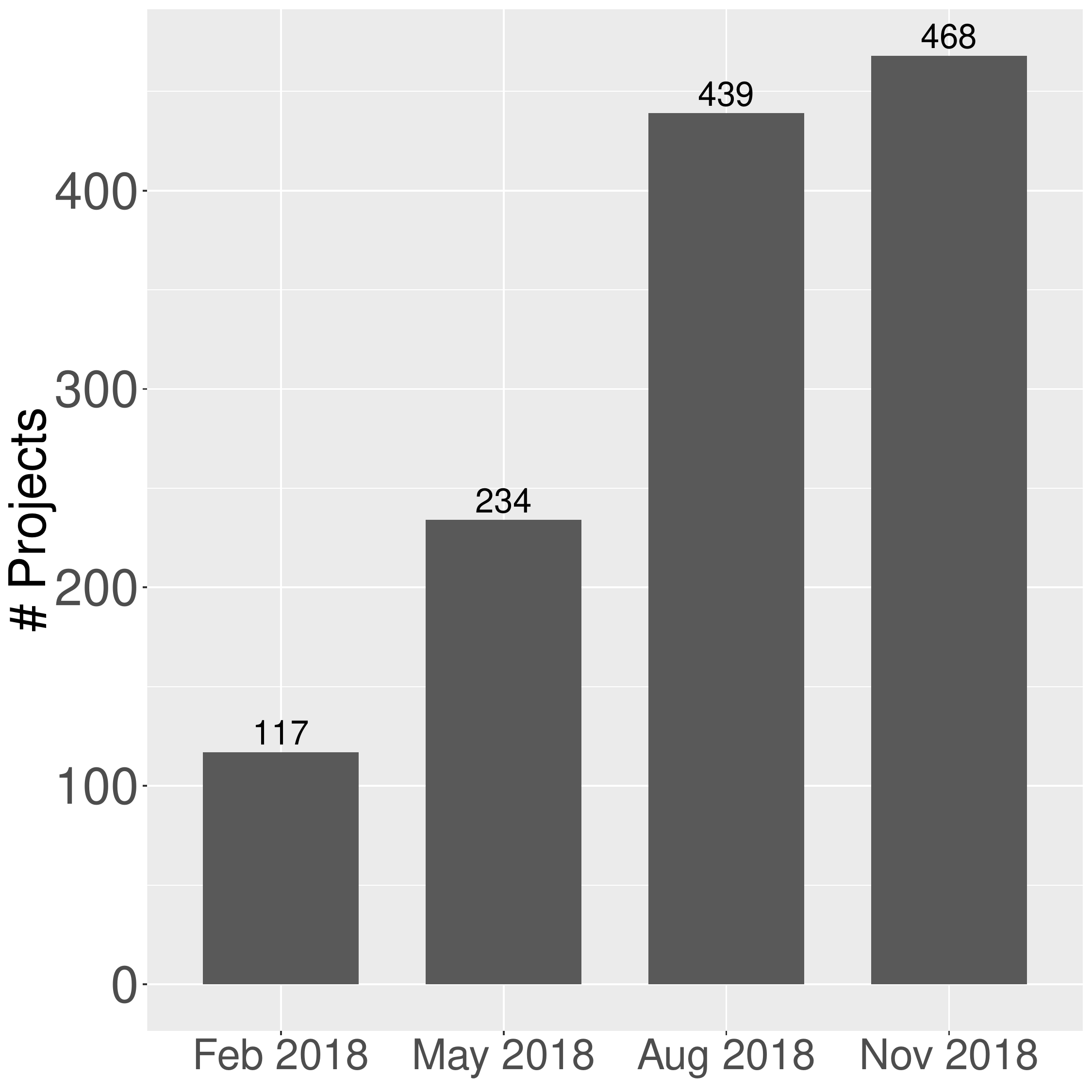}
\caption{Number of new unmaintained projects over time.}
\label{fig:projects-become-unmaintained-over-year}
\end{figure}

\subsection{Historical Analysis}
\label{sec:lma-historical-analysis}

In this section, we analyze the historical evolution of 2,927 active projects, as classified by our model in November, 2017. To build a trend line, we compute new LMA values for these projects in November, 2018, i.e.,~after one year. Our goal is to study how often projects become unmaintained and whether the LMA values change significantly over time. Particularly, we compute LMA values in intervals of 3 months during the period of analysis, i.e.,~February 2018, May 2018, August 2018, and November 2018. We also evaluate LMA values under two perspectives: programming language and application domain. Figure~\ref{fig:projects-become-unmaintained-over-year} shows the total number of projects classified as {\em unmaintained} in each time interval. As we can see, the number of unmaintained projects is increasing over time, moving from 117 projects (4\%) in February 2018 to 468 projects (16\%) in November 2018.

\begin{figure}[!ht]
\centering
\includegraphics[width=7cm]{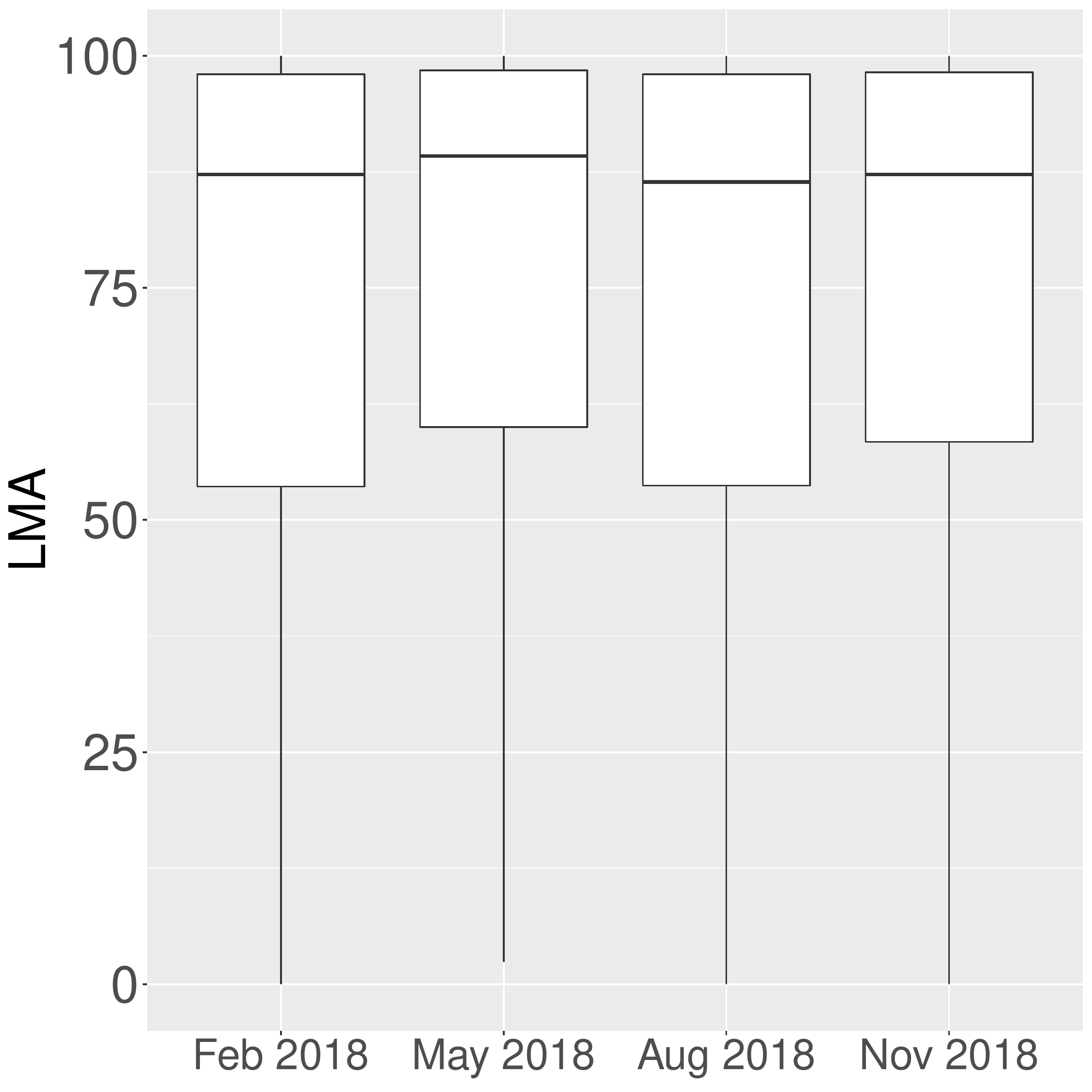}
\caption{LMA values over time.}
\label{fig:historical_lma}
\end{figure}

Figure~\ref{fig:historical_lma} shows the distribution of the LMA values on each period. The median values are 87.2, 89.2, 86.4, and 87.2, respectively. By applying Kruskal-Wallis to compare multiple samples, we found that these distributions are statistically different, but the difference is {\em negligible} by  Cliff's delta. Although there is an increasing number of unmaintained projects, there is also a significant number of projects with maximal LMA values. In each interval, we found 215, 230, 208, and 197 projects with maximal LMA values, respectively. Some popular projects---such as {\sc rails/rails}, {\sc matplotlib/matplotlib}, and {\sc numpy/numpy}---have maximal LMA in all considered time frames.

\begin{figure}[!t]
\centering
\includegraphics[width=12cm]{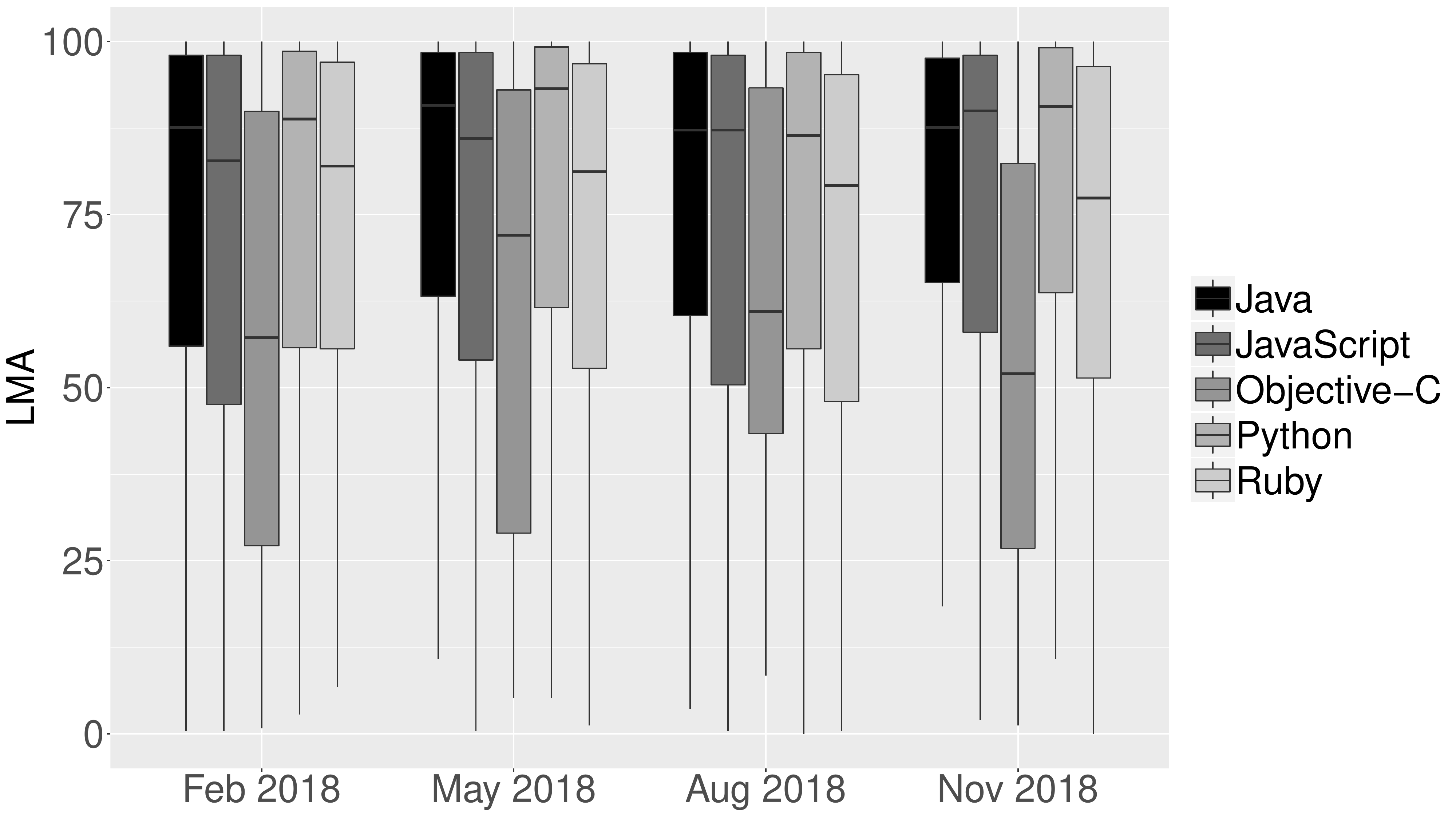}
\caption{Historical LMA values by programming language.}
\label{fig:historical_lma_by_language}
\end{figure}

Figure~\ref{fig:historical_lma_by_language} shows the LMA values by programming language over the studied 3-month time intervals. We consider only the top-5 languages by number of projects, which are JavaScript (700), Python (360), Java (259), Ruby (216), and Objective-C (179). For all languages, the LMA values remain stable throughout the studied intervals, with the exception of Objective-C. These projects increased their LMA values from 57.2 in February 2018 to 72.0 in May 2018, but then decreased to 61.0 in August 2018 and 52.0 in November 2018.

\begin{figure}[!ht]
\centering
\includegraphics[width=12cm]{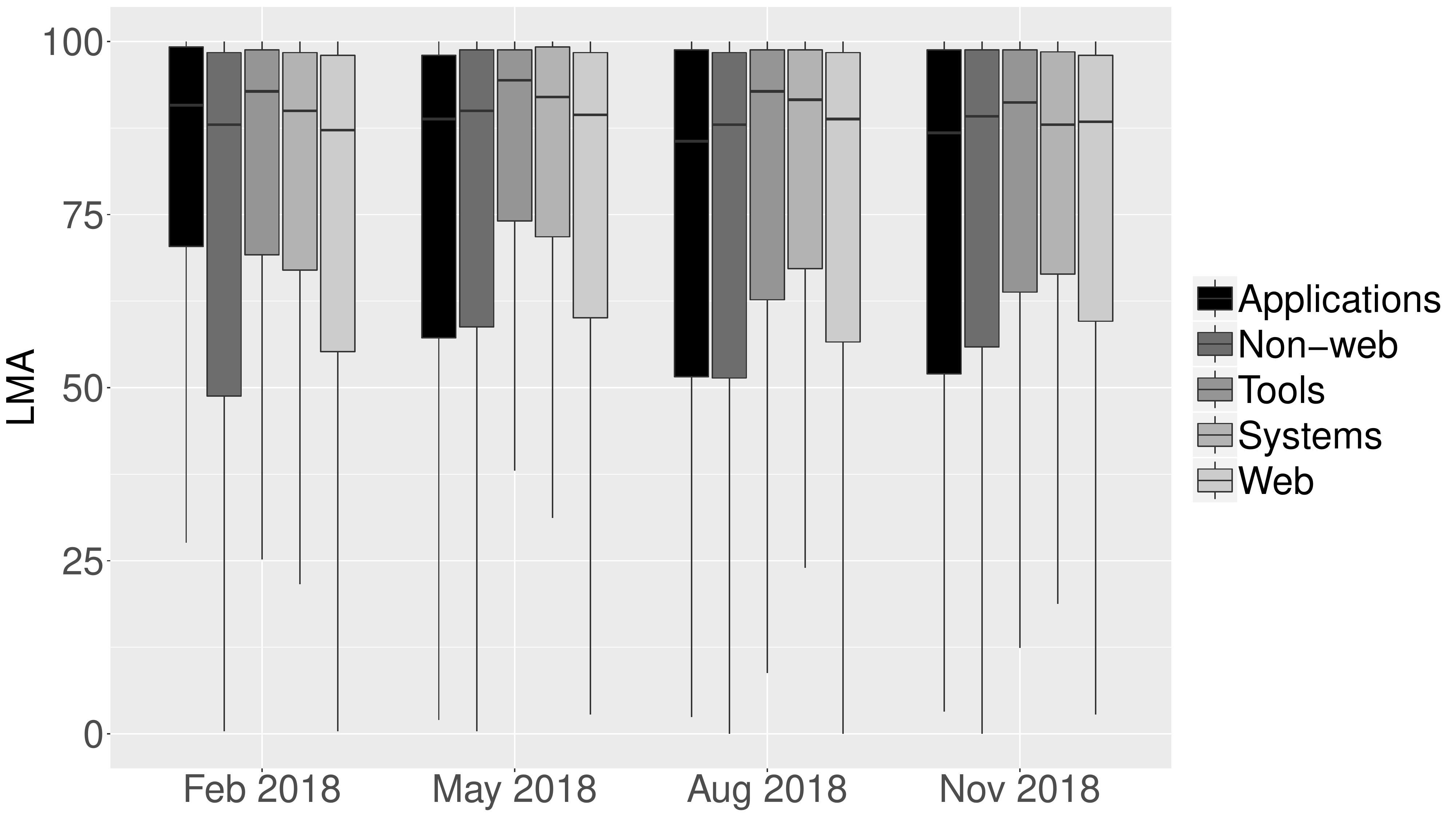}
\caption{Historical LMA values by application domain.}
\label{fig:historical_lma_by_domain}
\end{figure}

Finally, Figure~\ref{fig:historical_lma_by_domain} shows the historical LMA values by application domain. We use the same domains from the survival analysis (Section~\ref{sec:characteristics-of-unmaintained}). We found no significant differences between the distributions in the analyzed time frames. However, {\em Software Tools} have higher LMA values in all considered time intervals, which median measures 92.8, 94.4, 92.8, and 91.2, respectively.

\begin{formal}
From 2,927 active projects in November 2017, 468 projects (16\%) moved to an unmaintained state in the time interval of one year. We also found that Objective-C projects have lower LMA values than projects implemented in other programming languages. Finally, Software Tools have the highest LMA values over time.
\end{formal}

\subsection{Chrome Extension}

We implemented a Chrome extension called {\em isMaintained} to indicate whether a GitHub project is actively maintained or not. This extension is publicly available at the Chrome Store.\footnote{\url{https://chrome.google.com/webstore/search/ismaintained}} It only installs a small icon on the right side of a repository's page. This icon's color is used to inform the maintenance level of a project. The projects classified as {\em unmaintained} have a red icon. On the other hand, the level of maintenance activity for active projects can be high, fair, or borderline. The projects with LMA values in the fourth quartile of LMA values are labeled as high (green icon); projects in the second and third quartiles are labeled as fair (yellow icon); and projects in the first quartile are labeled as borderline (orange icon). Finally, the remaining repositories in our dataset (e.g.,~books, tutorials, awesome-lists, etc) receive a grey icon. Table~\ref{tab:lma_levels} shows the levels of maintenance activity and their respectively color used to classify the projects.

\begin{figure}[!ht]
\centering
\makebox[\textwidth]{\includegraphics[width=\textwidth]{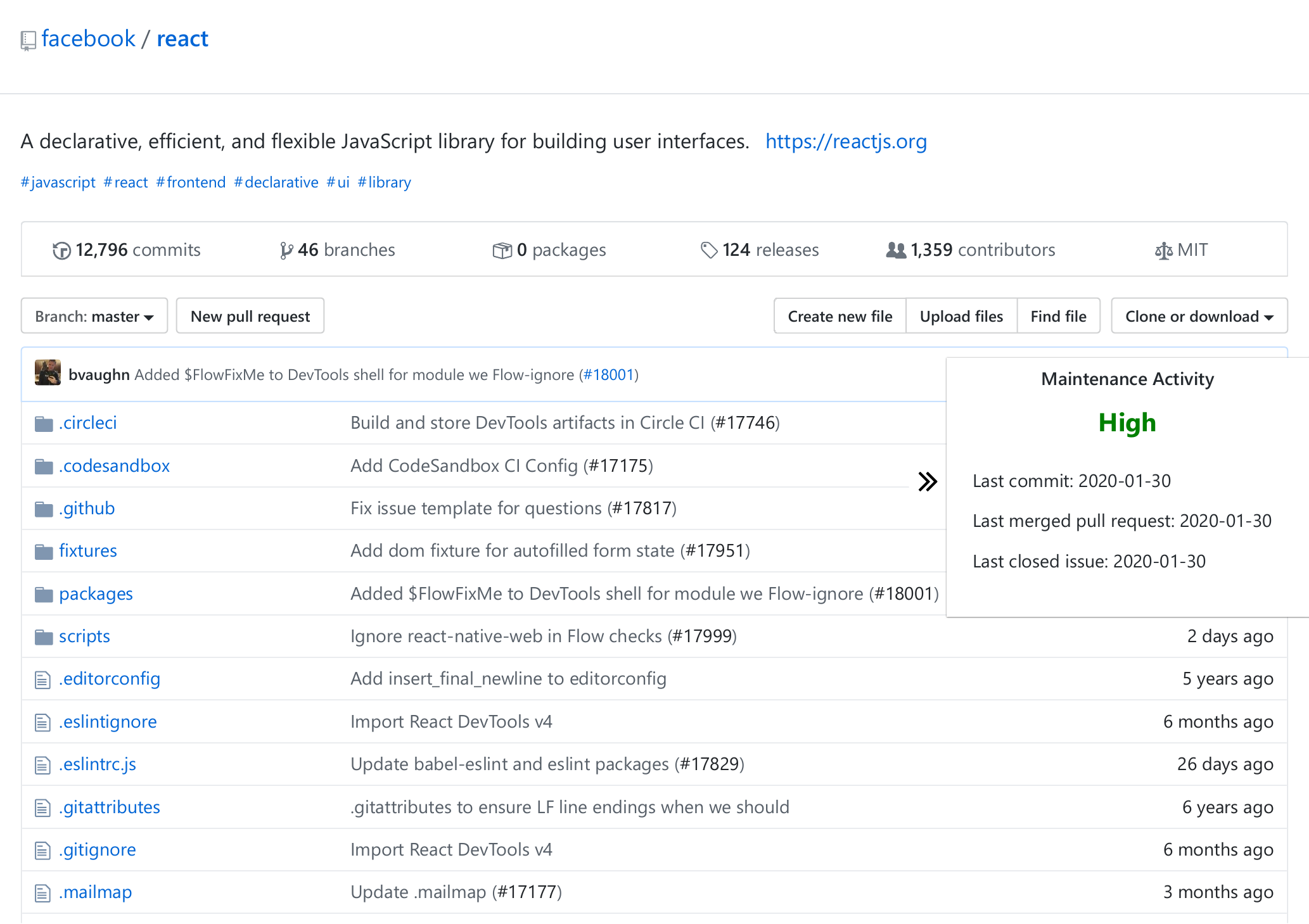}}
\caption{Example of a GitHub page using the LMA plugin (on the right side).}
\label{fig:example-githubpage-using-plugin}
\end{figure}

Figure~\ref{fig:example-githubpage-using-plugin} shows an example of a GitHub page with the proposed Chrome extension enable. In this example, we show the level of activity for {\sc facebook/react} with a green icon (high maintenance activity).

\begin{table}[!ht]
    \centering
    \caption{Levels of maintenance activity as considered by the implemented Chrome extension.}    
    \begin{tabular}{ l l l}
        \toprule
      {\bf Level of activity}    & {\bf Color} & {\bf LMA Quartile} \\ 
        \midrule
        High	   	    & green     & 4th   \\
        Fair	   	    & yellow    & 2nd and 3rd \\
        Borderline      & orange    & 1st   \\ 
        Unmaintained    & red       & -     \\
        Not analysed    & grey      & -     \\
        \bottomrule
    \end{tabular}
    \label{tab:lma_levels}
\end{table}

\subsection{Examples of reuse scenarios}
 
In this section, we show two examples on how the proposed model can help developers to make the decision of selecting a library on GitHub. We described the two scenarios as follows:\\[-.25cm]
 
\noindent{\bf First scenario:} Suppose that a developer is searching for a communication library to implement a client/server application. In the end of his/her searching, the developer ends up with a list of the following libraries: {\sc zaphoyd/websocketpp}, {\sc chriskohlhoff/asio}, {\sc grpc/grpc}, and {\sc scylladb/seastar}. The level of maintenance activity (LMA) of these libraries are Fair, Fair, High, and High, respectively. All these projects are popular on GitHub, since they have more than 2K stars. Their main programming language is C++. Based on the results of our survival analysis, only 30\% of the projects implemented in C++ survive more than 7 years. The age of these libraries are 8, 8, 5, and 6 years, respectively. By combining this information, we eliminate the projects {\sc zaphoyd/websocketpp} and {\sc chriskohlhoff/asio} due to their LMA values. Next, we realize that the survival probability of {\sc grpc/grpc} is higher than the one of {\sc scylladb/seastar}. Therefore, we consider that {\sc grpc/grpc} might the better choice. However, we clarify this is solely a quantitative and data-driven analysis. Other factors---such as the features provided by each library---should be considered as well.\\[-.25cm]
 
\noindent{\bf Second scenario:} Suppose that a developer is looking for a JavaScript testing framework. He/she ends up with a list of the following frameworks: {\sc cypress-io/cypress}, {\sc mochajs/mocha}, {\sc jas\-mine/jas\-mine}, and {\sc facebook/jest}. All these projects are very popular on GitHub with at least 5K stars. Following our model results, we found that the level of maintenance activity (LMA) of these projects are Fair, Fair, High, and High, respectively. Based on the survival analysis results, we found that only 10\% of the projects implemented in JavaScript survive more than 7 years. The age of these libraries are 5, 8, 11, and 5 years, respectively. By combining the two approaches, we remove the projects {\sc cypress-io/cypress} and {\sc mochajs/mocha} due to their lower LMA values. Therefore, we consider that {\sc facebook/jest} might be an interesting choice, since it is a younger framework.

\section{Threats to Validity}
\label{sec:threats}

The threats to validity of this work are described as follows:\\[-.25cm]

\noindent{\bf External Validity:} Our work examines open source projects on GitHub. We recognize that there are popular projects in other platforms (e.g., Bitbucket, SourceForge, and GitLab) or projects that have their own version control installations. Thus, our findings may not generalize to other open source or commercial systems. A second threat relates to the features we have considered. By adding other features, we may improve  the prediction of unmaintained projects; however, given our high prediction performance, we feel confident that our features are effective. Also, some of the features we use may not be available in other projects, however, most of our features are available in most code control repositories/ecosystems. In the future, we intend to investigate additional projects and consider more features.\\[-.25cm]

\noindent{\bf Internal Validity:} The first threat relates to the selection of the survey participants. We surveyed the project owner, in the case of repositories owned by individuals, or the developer with the highest number of commits, in the case of repositories owned by organizations. We believe the developers who replied to our survey are the most relevant given their level of activity in the project. It is also possible that most missing answers are from developers of unmaintained projects. As a second threat, the themes of the survey were defined and organized by the authors of the paper. As with any human activity, the derived themes may be subject to bias and different researchers might reach different observations. However, to mitigate this threat, a first choice of themes was conducted in parallel by the first two authors of this paper. Also, they attended several meetings during a whole week to improve the initial selected themes. A third threat relates to the parameters used to perform our experiment. We set the number of trees to 100 to train our classifier. To attenuate the bias of our results, we run 5-fold cross-validation and use the average performance for 100 rounds. A forth threat is related to how the accuracy of our machine learning approach was evaluated. We relied on developer replies about their projects to evaluate the performance of our machine learning classifier. In some cases, the developer replies (or developers who did not reply) may impact our results. That said, our survey had a response rate of 37.1\%, which is very high for a software engineering study, giving us confidence in the reported performance results.\\[-.25cm]

\noindent{\bf Construct Validity:} A first threat relates to the definition of active projects. We consider as active projects those with at least one release in the last month (Section~\ref{sec:machine-learning}). We acknowledge a threat in the definition of the time frame. To mitigate this threat, the first paper's author inspected each selected project to look for deprecated projects (\totalProjectsWithDeprecatedMessageInReadmeOfSurvey\ projects declare they are no longer being maintained) and we conduct a survey with \totalSurveyAnswersByDevelopers\ developers to confirm our findings. A second threat is related to the projects we studied. Our dataset is composed of the most starred projects (and additional filtering). Although the starred projects may not be representative of all open source projects, we did carefully select such projects to ensure that our study is conducted on real (and not toy) projects.  

\section{Related Work}
\label{sec:related-work}

\noindent{\bf Machine Learning.} 
Recently, the application of machine learning in software engineering contexts has gained much attention. Several researchers have used machine learning to accurately predict defects (e.g.~\cite{peters2013better}), improve issue integration (e.g.,~\cite{Alencar2014}), enhance software maintenance (e.g.,~\cite{gousios2014exploratory}), and examine developer turnover (e.g.,~\cite{bao2017will}). For example, \citet{gousios2014exploratory} investigate the use of machine learning to predict whether a pull request will be merged. They extract 12 features organized into three dimensions: pull request, project, and developer. They conduct their study using six algorithms (Logistic Regression, Naive Bayes, Decision Trees, AdaBoost with Decision Trees, and Random Forest). \citet{bao2017will} build a model to predict developer turnover, i.e., whether a developer will leave the company after a period of time. They collect several features based on developers monthly report from two companies. The authors evaluate the performance of five classifiers (KNN, Naive Bayes, SVM, Decision Trees, and Random Forest). In both studies, Random Forest outperforms the results of other algorithms. In another study, \citet{martin2016causal} train a Bayesian model  to support app developers on causal impact analysis of releases. They mine time-series data about Google Play app over a period of 12 months and survey developers of significant releases to check their results.
\citet{tian2015characteristics} use Random Forest to predict whether an app will be high-rated. They extract 28 factors from eight dimensions, such as app size and library quality. Their findings show that external factors (e.g., number of promotional images) are the most influential factors. Our study also uses machine learning techniques, however, our main goal is to detect projects that are not going to be actively maintained. Moreover, our study extracts project, contributor and owner features that we input to the machine learning models.\\[-.3cm]

\noindent{\bf Open source projects maintainability.} In previous work~\citep{coelho2017why}, we survey maintainers of 104 failed open source projects to understand the rationale for such failures. Their findings revealed that projects fail due to reasons associated with project properties (e.g., low maintainability), project team (e.g., lack of time of the main contributor), and to environment reasons (e.g., project was usurped by a competitor). Later, we report results of a survey with 52 developers who recently became core contributors on popular GitHub projects~\citep{coelho2018why}. Our results show the developer's motivations to assume an important role in FLOSS projects (e.g., to improve the projects because they use them), the project characteristics (e.g., a friendly community), and the obstacles they faced (e.g., lack of time of the project leaders). 

Also related is the work by \citet{yamashita2014magnet}, which adapts two population migration metrics in the context of open source projects. Their analysis enables the detection of projects that may become obsolete.  \citet{khondhu2013all} report that more than 10,000 projects are inactive on SourceForge. They use the maintainability index (MI)~\cite{oman1992metrics} to compare the maintainability between inactive projects and projects with different statuses (active and dormant). Their results reveal that the majority of inactive systems are abandoned with a similar or increased maintainability, when compared to their initial status. Nonetheless, there are critical concerns on using MI as a predictor of maintainability~\cite{bijlsma2012faster}. \citet{nadia2016roads} reports risks and challenges to maintain modern open source projects. She argues that open source plays a key role in the digital infrastructure of our society today. Opposed to physical infrastructure (e.g., bridges and roads), open source projects still lack a reliable and sustainable source of funding.

\citet{liu2018recommending} present a learning-to-rank model to recommend open source projects for developers. \citet{rastogi2018relationship} investigate 70,000+ pull requests from 17 countries to model the relationship between the geographical location of developers and pull request acceptance decision. \citet{steinmacher2018almost} conducted surveys with quasi-contributors to understand their perceptions for pull-request non-acceptance. Their results show that non-acceptance discourage developers to submit new pull-requests. \citet{barcomb2019episodic} show five factors that affect retention of episodic volunteers in FLOSS communities. Other recent research on open source has focused on the organization of successful open source projects~\cite{mockus2002two} and on how to attract and retain contributors~\cite{zhou2015will, steinmacher2016overcoming, leeunderstanding, pinto2016more, canfora2012going}.\\[-.3cm]

\noindent{\bf Survival analysis.} Survival analysis was first used in the medical domain and then applied to other domains including software engineering. For example, Maldonado et al.~\cite{maldonado2017empirical} use survival analysis to determine how long self-admitted technical debt lives in a project before it is actually removed. \citet{lin2017developer} applied survival analysis on five open source projects to understand the impact of several factors on developers leaving a project. \citet{valiev2018ecosystem} use survival analysis on a large set of PyPI projects hosted on GitHub. \citet{samoladas2010survival} proposed a framework for assessing the survival probability of a FLOSS project and evaluate the benefits of adding more committers in a project. \citet{businge2012survival} investigate the survival of 467 third-party Eclipse plug-ins. Different from this works, we use survival analysis to reveal the survivability probability of a large scale of open source projects under different perspectives (e.g.,~organizational or individual account, programming language, and application domain).

\section{Conclusion}
\label{sec:conclusion}

In this paper, we proposed a machine learning model to identify unmaintained GitHub projects and to measure the level of maintenance activity (LMA) of active GitHub projects. By our definition, the {\em unmaintained} status includes three types of projects: finished projects, deprecated projects, and stalled projects. We validated the proposed model with the principal developers of \totalSurveyAnswersAndReadme\ projects, achieving a precision of \modelPrecision\% (RQ1). Then, we used the model with \totalProjectsUnmaintainedByReadme\ deprecated projects---as explicitly mentioned in their GitHub page. In this case, we achieved a recall of \modelRecall\% (RQ2). We also showed that the proposed model can identify unmaintained projects early, without having to wait for one year of inactivity, as commonly proposed in the literature (RQ3). We assessed the survival probability of unmaintained projects under three perspectives: organizational or individual account, programming language, and application domain (RQ4). We found a negligible difference on the survival probabilities of projects owned by individual and organizational accounts. Moreover, Ruby projects have higher probabilities of survival. Regarding the analysis by application domain, we found that {\em System Software} is the domain with the highest survival probability. Finally, we investigate whether unmaintained projects follow (or not) a set of best open source contribution practices (RQ5). Our results show that the practices with the highest effect are continuous integration, followed by the adoption of contributing guidelines, and the presence of labels to recommend issues to newcomers.

Finally, we defined a metric, called Level of Maintenance Activity (LMA), to assess the risks of projects become unmaintained. We provided evidence on the applicability of this metric by investigating its usage in \testsetClassifiedAsActive\ projects classified as active in our dataset. We evaluate the LMA of these projects in the time frame of one year under two different perspectives: programming language and application domain. We found that 16\% become unmaintained over this time. We also reported that Objective-C projects have lower LMA values than projects implemented in other languages. Software Tools have the highest LMA values over time. Finally, we implemented a public Chrome extension called {\em isMaintained} to show the level of maintenance activity of a GitHub project. This extension is publicly available at: \url{https://chrome.google.com/webstore/search/ismaintained}.

As future work, we intend to improve our Chrome extension to automatically mine and evaluate new GitHub projects and calculate their LMA every three months. 

The dataset used in this paper is available at: \url{https://zenodo.org/record/1313637}.

\section*{Acknowledgments}
\noindent{Our research is supported by CAPES, FAPEMIG, and CNPq. We would also like to thank the \totalSurveyAnswersByDevelopers\ GitHub developers who kindly answered our survey.}

\balance
\bibliographystyle{Reference-Format}
\bibliography{ist_2020}

\end{document}